\documentclass[referee,sn-apa]{sn-jnl} 

\usepackage{amsmath, graphicx, tabularx, longtable, booktabs, threeparttable, caption, enumitem, url, lmodern}

\usepackage[T1]{fontenc}

\usepackage{hyperref}

\makeatletter
  {\begingroup\ttfamily\catcode`\_=12 }%
  {\endgroup}
\makeatother

\begin{document}
    
\title{Personalized Multimodal Feedback Using Multiple External Representations: Strategy Profiles and Learning in High School Physics}

\author*[1]{Natalia Revenga-Lozano}
\email{natalia.revenga@lmu.de}
\author[1]{Karina E. Avila}
\author[1]{Steffen Steinert}
\author[1]{Matthias Schweinberger} 
\author[1]{Clara E. Gómez-Pérez}
\author[1]{Jochen Kuhn}
\author*[1]{Stefan Küchemann}
\email{s.kuechemann@lmu.de}

\affil[1]{Chair of Physics Education, Faculty of Physics, Ludwig-Maximilians-Universität München (LMU Munich), Geschwister-Scholl-Platz 1, 80539 Munich, Germany}

\abstract{
Multiple external representations (MERs) and personalized feedback support physics learning, yet evidence on how personalized feedback can effectively integrate MERs remains limited. This question is particularly timely given the emergence of multimodal large language models. We conducted a 16-24 week observational study in high school physics (N=661) using a computer-based platform that provided verification and optional elaborated feedback in verbal, graphical and mathematical forms. Linear mixed-effects models and strategy-cluster analyses (ANCOVA-adjusted comparisons) tested associations between feedback use and post-test performance and moderation by representational competence. Elaborated multirepresentational feedback showed a small but consistent positive association with post-test scores independent of prior knowledge and confidence. Learners adopted distinct representation-selection strategies; among students with lower representational competence, using a diverse set of representations related to higher learning, whereas this advantage diminished as competence increased. These findings motivate adaptive feedback designs and inform intelligent tutoring systems capable of tailoring feedback elaboration and representational format to learner profiles, advancing personalized instruction in physics education.}


%\maketitle must follow title, authors, abstract, and keywords
\maketitle

% body of paper here - Use proper section commands
% References should be done using the \citep, \ref, and \label commands
\section{Introduction}
\label{sec: intro}

One of the main challenges in education is to be able to adapt instruction to the learner \citep{hardy_ilonca_adaptive_2019}. The value of adaptive learning in education is widely recognized with large empirical support: research on aptitude-treatment interactions (ATI) has consistently shown that instructional effectiveness varies depending on individual learner characteristics, such as prior knowledge, cognitive style, and motivation \citep{cronbach_aptitudes_1977, liu_investigating_2017}. In other words, different students benefit from different types of instruction, a reality that challenges the traditional "one-size-fits-all" approach to teaching. Various meta-analyses confirm the superiority of personalized instruction over uniform instructional methods \citep{deunk_differentiation_2015, kulik_effectiveness_1990}.

Despite this strong theoretical and empirical basis, actual implementation of personalized learning in classroom settings remains limited. A major constraint lies in the impracticality of one-to-one instruction in typical educational environments, where limited time, lack of teacher support, and rigid curriculum structures hinder customized instruction \citep{aleven_instruction_2016}.

In this context, technology-enhanced learning environments offer a promising avenue. Digital tools can provide flexible infrastructures to deliver personalized instruction that dynamically adapts to the individual needs of learners. Meta-analyses of computer-based learning environments have demonstrated their overall effectiveness compared to traditional instruction, particularly when the learning environment is adaptive \citep{cheung_effectiveness_2013, tamim_what_2011}: environments that adjust timing, content, modality or feedback based on learner behavior tend to produce significantly better outcomes \citep{aleven_instruction_2016, faber_effects_2017}.

However, despite the growing body of evidence supporting the use of adaptive technology, fundamental questions remain open: when should adaptation occur? How should it be implemented? And which specific adaptable components of a digital learning environment are the most effective? \citep{durlach_adaptive_2012, faber_effects_2017}. A very powerful tool for instruction and, in particular, for individualized instruction, is feedback. Therefore, insights into how to adapt feedback are especially relevant for research on adaptive technologies \citep{van_der_kleij_effects_2015}. Additionally, in the context of physics education, another important area subjected to adaptability is the use of multiple external representations (MERs) or multimodalities (e.g., graphs, diagrams, equations) \citep{opfermann_multiple_2017}. Consequently, shedding light on what strategies for the use of MERs can be potentially most effective with different students can contribute to the design of effective adaptive systems.

Last but not least, with the advent of Artificial Intelligence (AI), and specifically, of Large Language Models (LLMs) \citep{vaswani_attention_2017} and Multimodal Large Language Models (MLLMs) \citep{gemini_team_gemini_2023}, new possibilities for adaptive systems emerge in the form of intelligent tutoring systems (ITSs) that can provide learners with immediate, individualized feedback, while controlling the scope and style of the feedback \citep{gaeta_enhancing_2025, olney_enhancing_2024, yan_promises_2024}. Real-time AI-driven feedback within ITSs is associated with learning gains and faster learning in several studies, although advantages over non-intelligent systems are mixed \citep{letourneau_systematic_2025}. In addition to real-feedback, especially in the field of physsics, it is crucial to be able to create instructional material that includes multimodalities or multiple representations (MERs) (e.g. diagrams, graphs, equations). MERs provide complementary information and disciplinary advantages, and learning to coordinate and translate among them improves understanding and problem solving \citep{ainsworth_deft_2006, kohl_understanding_2017, fredlund_unpacking_2014}. Therefore, it is important that MLLMs are incorporated into ITSs in such a way that they provide correct, specific MERs in the amount and timing that are appropriate for the student. A thorough investigation of effective strategies for designing instruction with MERs is relevant for an optimal implementation of MLLMs in physics education \citep{bewersdorff_taking_2025}.

\subsection{Feedback}
\label{subsec:feed}
Feedback is known to be one of the most effective interventions to foster student learning \citep{hattie_instruction_2011, hattie_power_2007}. A recent meta-analysis on the topic has reported a medium effect of $d=0.48$ on feedback on student learning \citep{wisniewski_power_2020}. A theoretical explanation for the positive effect of feedback on learning outcomes is given by \citep{butler_feedback_1995}, who argue that feedback is information with which the learner can confirm, add to, overwrite, tune, or restructure information in memory. This helps students identify and correct errors and misconceptions, develop more effective and efficient problem-solving strategies, and improve their self-regulation, provided that feedback is processed with care \citep{bangert-drowns_instructional_1991}. However, the literature shows that the variations are large with regard to the effect of feedback when different aspects of instruction, student characteristics, learning goals, and educational contexts are considered, ranging from negative to large effects \citep{wisniewski_power_2020, van_der_kleij_effects_2015, shute_focus_2008}. 

Having a good understanding of the different classifications of feedback and the variables that can moderate its effect is essential for informing quality research and the design of feedback strategies in different educational contexts. For example, feedback can be given at different levels of cognitive complexity, to address different learning goals, with different levels of elaboration, through different channels, and at different timings \citep{hattie_power_2007}.

A central design choice is the level of elaboration: from verification feedback (stating correct/incorrect) to elaborated feedback that addresses the topic, addresses the response, discusses specific errors, provides worked examples, or offers gentle guidance \citep{shute_focus_2008}. Literature findings point towards a certain consensus in this regard: the more relevant information is given, the better. This means that verification feedback effects are, in general, smaller than elaborated feedback effects. Especially effective are elaborated forms of error modeling (e.g. explaining why a learner made a mistake and how to avoid it, rejecting erroneous hypotheses, or suggesting effective strategies). It is also important that the feedback is non-evaluative, non-controlling, timely, brief, and specific, especially for task-level feedback \citep{wisniewski_power_2020, shute_adaptive_2008}. 

There are several factors to take into account when deciding what type of feedback is optimum. For example, systematic reviews report that verification feedback can be as effective, or even superior, depending on task complexity, prior knowledge, response certitude, time constraints, and performance goals \citep{shute_adaptive_2008, van_der_kleij_effects_2015, kluger_effects_1996}. Therefore, determining which level of elaboration benefits a given student, given their characteristics and the context of the task, is a key step toward building an ITS.

\subsection{Multiple External Representations in Physics Learning}
\label{subsec: mers}

The use of MERs in physics is crucial for mastering this science for two main reasons. First, complex natural phenomena often require more than one type of representation to be fully understood, processed, and communicated \citep{stylianou_problem_2020, johri_role_2013, opfermann_multiple_2017}. For instance, verbal representations, whether spoken or written, can express highly abstract, nuanced relationships, but may become cumbersome or lack the precision needed in exact sciences. In contrast, images, diagrams, and sketches can more intuitively depict the natural and often dynamic phenomena, though they may fall short in capturing motion or change. Graphs, for example, better represent dynamic relationships between variables but add a layer of abstraction. Finally, natural phenomena relies heavily on mathematical modeling to describe causal phenomena and to generate quantitative predictions that can be experimentally tested. However, mathematical modeling poses challenges for learners who struggle to connect mathematical expressions with their physical meaning \citep{angell_physics_2004}.

Extensive empirical evidence shows that the use of MERs can facilitate learning and problem-solving skills across educational contexts and subjects, promoting deeper conceptual understanding by highlighting complementary aspects of the content \citep{ainsworth_deft_2006}. The classical information processing approach, such as the Cognitive Theory of Multimedia Learning (CTML) or the Integrated model of Text and Picture Comprehension (ITPC), explains this by arguing that working with MERs produces a more efficient and effective use of cognitive resources \citep{mayer_cambridge_2005, schnotz_integrated_2005}. Other theories, such as the model of distributed and embodied cognition, argue for a much more interactive relation between the learner and the MERs \citep{pande_representational_2017}: external representations interact with internal ones, enabling engagement with phenomena otherwise imperceptible. They play an essential role in changing cognition, generating ideas, and activating imagination \citep{hutchins_distributed_2000}. In the words of D. Kirsh, \textit{external representations allow us to think the previously unthinkable}  \citep{kirsh_thinking_2010}. Additionally, embodied cognition emphasizes that not only neural processes but also sensorimotor processes are involved in the cognitive processing of MERs \citep{clark_intrinsic_2005, friston_embodied_2011}. 

Despite the extensive literature on the potential benefits of MERs, the actual processes by which learners select and interact with representations remain underexplored. In particular, little is known about how learners spontaneously engage with different types of representations when given freedom of choice. According to \citep{rexigel_more_2024}, the ability to choose an appropriate representation when presented with MERs may be an important advantage, as it allows learners to focus on relevant information and filter out redundant details. Despite the advantages of free choice of MERs, one of the most common difficulties for students when using MERs is the ability to switch from one representation to the other (known as representational fluency) \citep{chiou_study_2010, ibrahim_role_2013, bollen_student_2017}. Indeed, the study of \citep{schwonke_how_2009} indicates that learners rarely understand or engage with the intended functions of representations unless explicitly instructed. This suggests that representational competence does not emerge automatically, even in rich multimedia environments. Nonetheless, allowing learners some freedom in how they use multiple representations may still reveal important patterns in how they interpret, prioritize, or ignore different sources of information. These patterns, in turn, can offer valuable insights into how students navigate the representational landscape and where they may need additional support. Therefore, investigating how learners select and interact with representations in open-ended settings remains a critical step toward understanding the development of representational competence.

\subsection{Research questions}
\label{subsec:goals}

To inform the design of adaptive online instruction, we examine the roles of verification and elaborated feedback. Using prior-knowledge measures and interaction logs, we estimate their effects on performance in high-school physics.

We adopt an exploratory approach aimed at describing learners' selection patterns when interacting with verification and elaborated feedback delivered using MERs. We aim to observe and characterize how learners interact with representations and how these behaviors transfer to performance, while considering the interaction of prior knowledge and log-data. This approach may uncover patterns that can inform future experimental designs or targeted instructional interventions.

This paper addresses six research questions:
\begin{enumerate}
    \item[(RQ1):]Does the use of elaborated feedback have correlate positively with post-test scores when the alternative is verification feedback?
    \item[(RQ2):]Is there an interaction between the effect of elaborated feedback and other student characteristics like prior-knowledge, ability, or confidence?
    \item[(RQ3):]What are the MER(s)-selection patterns followed by the students?
    \item[(RQ4):]How do the MER-patterns that student follow relate to their performance in the post-test?
    \item[(RQ5):]How do the MER-patterns that student follow relate to their initial representational competences? 
    \item[(RQ6):]Do initial competence profiles moderate the effect of feedback strategies?
\end{enumerate}

\section{Methods}
\label{sec: methods}

We conducted a prospective observational Cohort study following high-school students over approximately one term to examine the naturally occurring use of elaborated feedback, presented in three different representational formats and as a complement to verification feedback. We additionally measured other individual characteristics such as response correctness, response certitude, initial representational competences, prior knowledge, and time used to solve each question. Our goal was to examine associations between these variables and the subsequent post-test performance.

\subsection{Participants and data collection}
To collect the data for this study we used a custom developed web application (KI4SCool) based on the open-source code from \url{https://gitlab.rhrk.uni-kl.de/ki4tuk/ratsapp} published by \citep{steinert_lessons_2025}. All materials related to the study were provided to the students through the web application: exercises, feedback, and assessment tests. The web application had an authentication architecture so that each student had unique credentials to access the platform from any location and device, and a unique anonymous \textit{id} associated with them. We worked with students enrolled in the 10th and 11th grades in high schools across Germany. A total of 29 schools, 87 classes, 1348 students from four federal states (Bayern, Hamburg, Rheinland-Pfalz and Hessen) were eligible for the study. Only 661 (51\% of total eligible) students from 66 classes and 26 schools finished pre- and post-tests. No a-priori power analysis was performed. The study was carried out over a period of 16 to 24 weeks, depending on the school, during the years 2023 and 2024. To examine whether attrition was systematic, in Appendix~\ref{app:attrition} we compared pre-test scores between participants who completed the study and those who dropped out.

\subsection{Materials}
\subsubsection{Pre-/Post-test and Exercise Sheets}
Before the kinematics instruction began at each school, we administered a pre-test to the students, with an identical post-test conducted after the lessons concluded (16 to 24 weeks later depending on the class and school). Only data from students who completed both tests is included in this study. We used two types of pre-/post-tests, with two different student Cohorts and two different approaches. 

For Cohort~1 (351 students from 34 classes and 10 schools from Bayern, Hamburg and Rheinland-Pfalz), we used a pre-test consisting of 9 questions designed in-house to measure both conceptual knowledge and representational competence simultaneously. These questions targeted three kinematic concepts (positive velocity, negative velocity, and acceleration) and repeated the same question using exclusively one of the external representations (textual, graphical, and mathematical) to provide the information necessary to solve the question. In this way, we decoupled the measurement of conceptual knowledge from representational competence, which allowed us to answer RQ5 and RQ6. However, the use of tests specifically designed for a concrete study is often criticized for the lack of generalization and the potential introduction of idiosyncrasies \citep{anglin_primer_2024}.

Thus, for Cohort 2 (310 students from 32 classes and 16 schools from Hessen), we used a 15-question test built with questions from validated tests to measure concept knowledge in kinematics, namely TUG-K \citep{beichner_testing_1994} (10 questions), FCI \citep{hestenes_force_1992} (3 questions), KiRC \citep{klein_assessment_2017} (1 question), and MBT \citep{hestenes_mechanics_1992} (1 question). The reason to use different questions from different tests was to cover what is taught in the school year without asking extra material that students were not supposed to be yet familiar with. The use of different assessment materials for Cohort 2 reduced our statistical power to answer RQ5 and RQ6, but enabled more robust results for the other research questions, as our conclusions were based on the results of two different post-tests.

After the study, we obtained a reliability index of
$r_{\text{test\_1}} = 0.77$ and $r_{\text{test\_2}} = 0.79$,
and Ferguson's delta of $\delta_{\text{test\_1}} = 0.97$ and
$\delta_{\text{test\_2}} = 0.98$, respectively, for each post-test, which, according to \citep{ding_approaches_2009}, corresponds to good levels of discrimination and internal consistency. The complete psychometric data for the pre-test/post-test is shown in Appendix~\ref{app:psychometrics}.

\begin{figure}[!htbp]
    \centering
    \includegraphics[width=13.5cm]{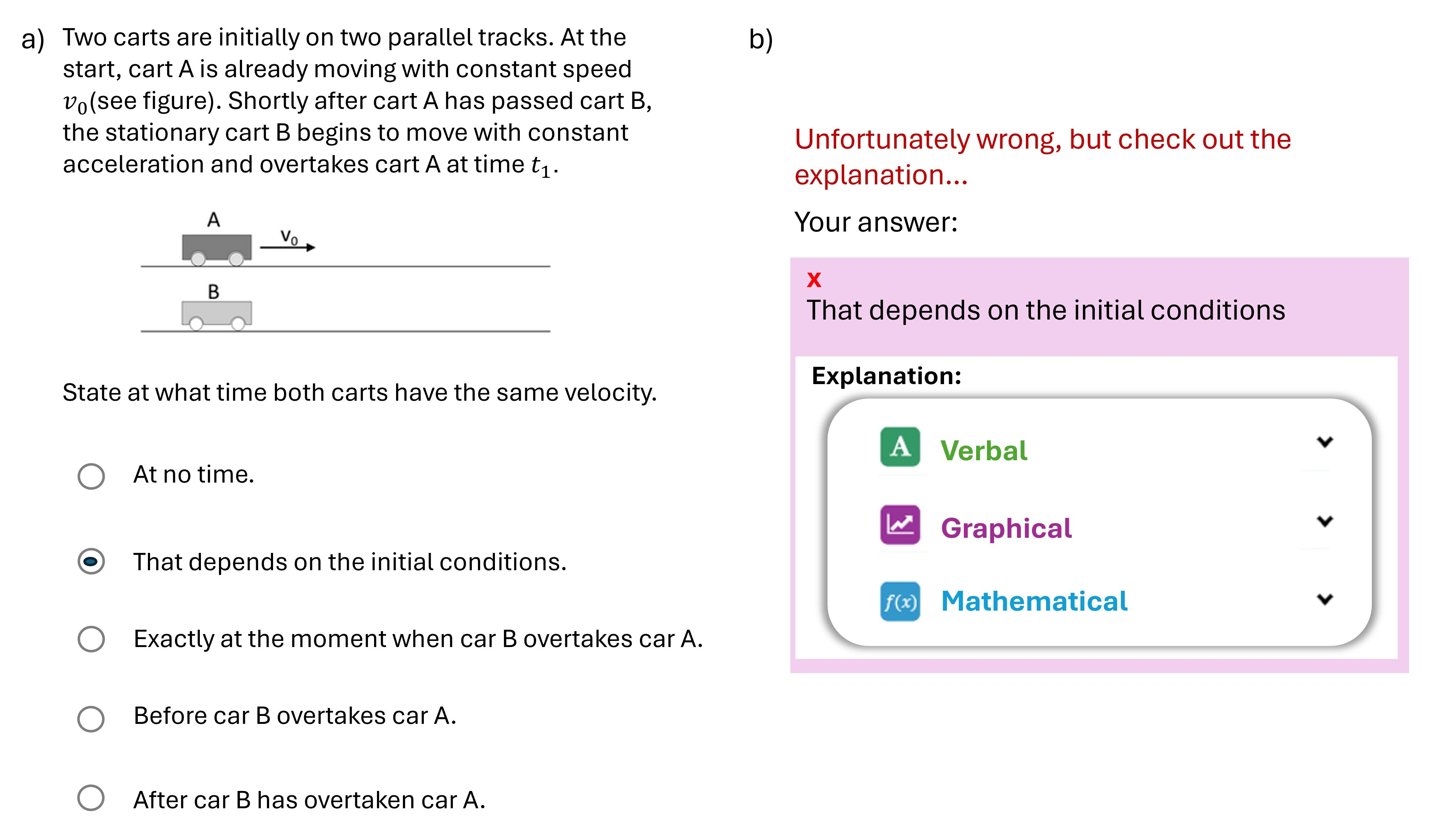}
    \caption{\textbf{Examples of physics exercises and automated feedback in the KI4SCool platform.} a) Example of a physics exercise in the platform KI4SCool, including the problem statement and the multiple-choice response options. b) Example of feedback displayed to students immediately after selecting an answer, including verification feedback and the option to get additional guidance through three different formats of elaborated feedback.}
    \label{fig:ex_and_feed_ex}
\end{figure}

The rest of the experiment continues as follows: after a certain kinematics topic was covered in class, we released an exercise sheet through the KI4SCool platform that contained 4-5 exercises related to the topic. The exercises were designed as short single-best-answer multiple-choice questions (SBA-MCQs) covering topics such as uniform linear movement, accelerated linear movement, free fall, and two-dimensional parabolic movement. In total, there were 43 exercises available. The exercises served as complementary practice material for the class. Students answered them primarily at home on a voluntary basis using their individual logins. See Fig.~\ref{fig:ex_and_feed_ex}~a) for an example of an exercise. We included in our analyses the data from all students who completed pre- and post-tests and solved at least 1 exercise in the platform (661 students), while controlling for this variables when analyzing our results.

\subsubsection{Feedback Presentation}

\begin{figure}[!htbp]
    \centering
    \includegraphics[width=14.5cm]{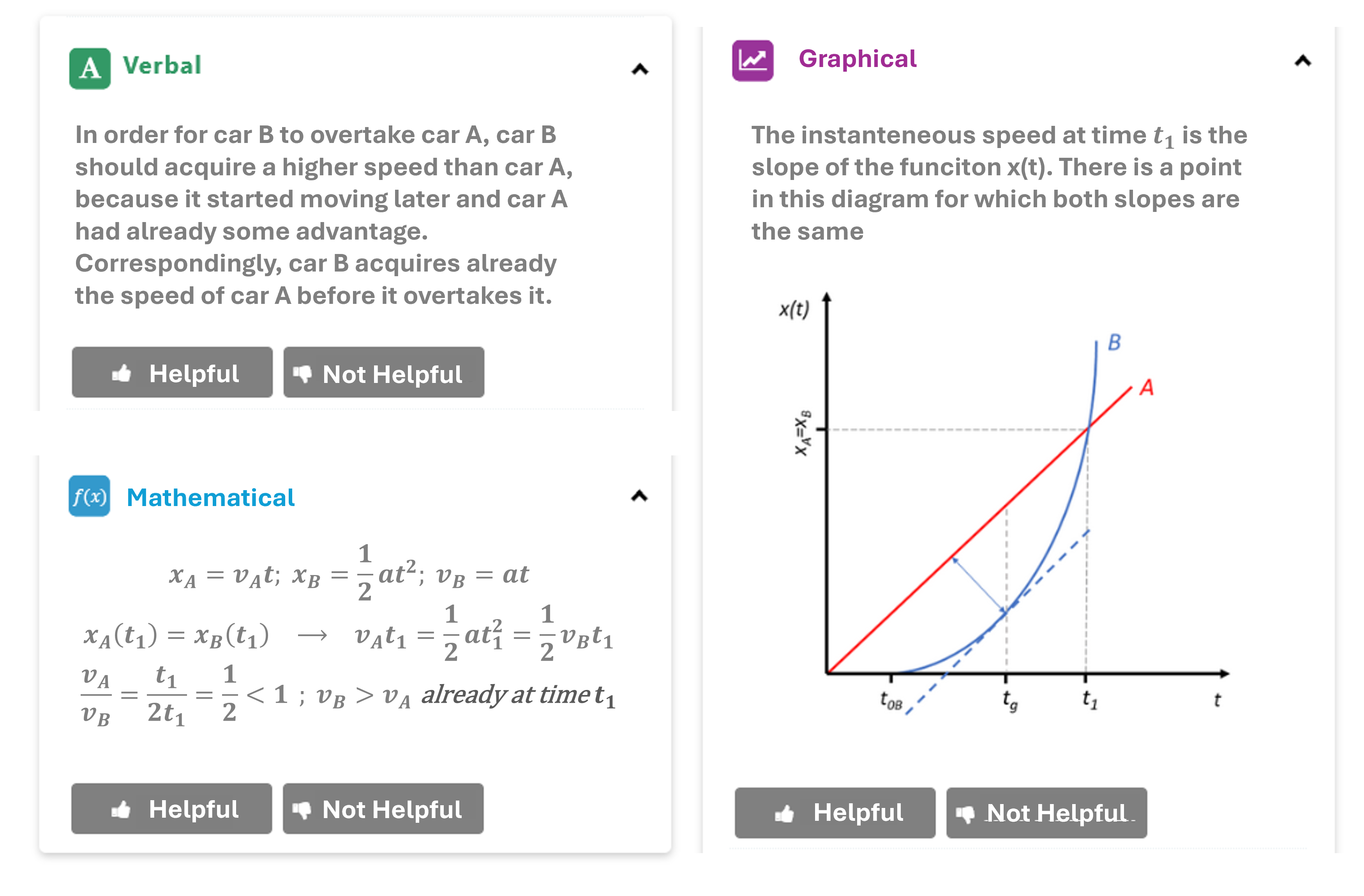}
    \caption{\textbf{Example of verification and elaborated feedback in the three representational formats.} Each format is displayed to the student only upon selection.}
    \label{fig:feed_ex_en}
\end{figure}

After each exercise was answered students received verification feedback, which provided information about the correctness or incorrectness of their answer (see Fig.~\ref{fig:ex_and_feed_ex}~b)). The interaction between feedback and students is designed as follows: verification feedback is shown in the upper part of the screen: a red rectangle with a cross symbol for wrong answers, a green rectangle with a tick symbol for correct answers. Additionally, students had the option to consult elaborated feedback, which is presented in the format of a drop-down menu, at first closed. The students are able to click on the arrows on the left of the screen to unfold the different explanations.  The elaborated feedback was presented in three external representations: textual, pictorial or graphical, and mathematical. The order of the representations (graphical, mathematical, verbal) is reshuffled for each question. See Fig.~\ref{fig:feed_ex_en} for examples of textual, pictorial/graphical, and mathematical feedback that refer to the exercise in Fig.~\ref{fig:ex_and_feed_ex}. Students were free to select the feedback format(s) to view: they could choose to open one, two or all feedback types at a time. Before seeing the feedback, students were asked how confident they were in the answer they gave (response certitude), on a scale from 1 (just a guess) to 5 (very confident). Both the physics questions and all the elaborated feedback used in this study were generated by the authors of this paper.

\subsection{Analysis plan}
\label{subsect:analysis-plan}
We analyzed post-test scores using linear mixed-effects regression with a random intercept for class/teacher to account for clustering. We then performed unsupervised k-means clustering on feedback-use (time-aggregated and time-resolved) summaries to derive strategy groups, and compared groups using ANCOVA, controlling for pre-test score and other covariates.

\subsubsection{Measures and pre-processing}
\label{subsection: preprocessing}
The primary outcome was the post-test score. Primary exposures were (i) the frequency of elaborated feedback use, (ii) the distribution of representational formats (verbal, graphical, mathematical), and (iii) the distribution of initial representational competences. The frequency of elaborated feedback use (or simply feedback frequency) is defined as the percentage of answered questions in which any type of selected feedback was selected ($\frac{\text{clicks on any feedback}}{\text{questions answered}}$). The distribution of representational formats is calculated using three variables: verbal frequency, graphical frequency and mathematical frequency. These are three independent usage frequencies computed for each student and that are defined as the proportion of answered questions for which the student selected that particular representation type ($\text{verbal frequency}=\frac{\text{clicks on verbal}}{\text{answers with feedback}}$), $\text{graphical frequency}=\frac{\text{clicks on graphical}}{\text{answers with feedback}}$), $\text{ mathematical frequency}=\frac{\text{clicks on mathematical}}{\text{answers with feedback}}$). Because students could select several representation types for the same question, these three frequencies are not mutually exclusive and are not constrained to sum to 1; a student who consistently selects all three representation types can obtain values close to 1 on all three variables. Initial representational competences are divided in verbal, graphical and mathematical, and are defined as the scaled scores on the verbal, graphical and mathematical questions of the pre-test for Cohort~1, respectively.

From the results of the pre-test we computed, per student, pretest score, and  initial verbal, graphical and mathematical competences. From platform logs we computed, per student, the total number of exercises solved, the count and proportion of elaborated-feedback openings (feedback frequency), and the proportion of openings by representational format (verbal frequency, graphical frequency and mathematical frequency). Additionally, we also computed, per student, the average response confidence (avg. confidence), the average time spend on reading and answering an exercise (avg. response time), and the percentage of answers that were answered correctly (platform score). Continuous predictors were mean-centered and standardized to unit SD where noted (Std. $\beta$ coefficients). Additionally, each student was assigned a unique vector of labels representing their unique sequence of feedback selection among the following possibilities: ["NF" (no-feedback), "V" (verbal), "G" (graphical), "M" (mathematical), "VG" (verbal+graphical), "VM" (verbal+mathematical), "GM" (graphical+mathematical), "VGM" (all three types)]. For students of Cohort~1, initial verbal, graphical and mathematical competences were also computed.

\subsubsection{Linear mixed-effects regression: RQ1-RQ2.}
\label{subsect: LR}
We fitted linear mixed-effects models of post-test score with fixed effects for pre-test score, platform score, exercises solved, average response confidence, and feedback frequency. To account for classroom dependence, we included a random intercept for class. Standard errors were calculated with class-clustered variance estimators (small-sample correction). We also introduced an interaction term between the platform score and the number of exercises solved after centering these variables. We applied this analysis to our both Cohorts independently. Models were fit in R with \texttt{lme4::lmer} \citep{bates_parsimonious_2015} and $p$-values for fixed effects used Satterthwaite degrees of freedom via \texttt{lmerTest} \citep{kuznetsova_lmertest_2017}. We report standardized coefficients and 95\%~CIs from \texttt{parameters::standardize\_parameters} \citep{ludecke_extracting_2020}. Marginal and conditional $R^2$ were computed with \texttt{performance::r2} \citep{ludecke_performance_2021}. 

Model form:
\begin{equation}
\begin{aligned}
y_{ij} &= \beta_0
  + \beta_1\,\text{(feedback frequency)}_{ij}
  + \beta_2\,\text{(pre-test score)}_{ij}
  + \beta_3\,\text{(platform score)}_{ij} \\
&\quad
  + \beta_4\,\text{(num.\ of exercises)}_{ij}
  + \beta_5\,\text{(avg.\ confidence)}_{ij}
  + u_j + \varepsilon_{ij}
\end{aligned}
\end{equation}
with $u_j \sim \mathcal{N}(0,\sigma_u^2)$ for class $j$ and $\varepsilon_{ij} \sim \mathcal{N}(0,\sigma^2)$.

School (three-level cluster) was not included as a third level because only 10 schools were available, which is insufficient for reliable variance estimation \citep{van_de_schoot_small_2020, maas_sufficient_2005}. %. and preliminary checks suggested negligible between-school variation... (I can do the preliminary test with hessen because I know well the schools they come from)
Following guidance to keep random-effects structures parsimonious and supported by the data, we retained random intercepts only. With small within-group n, maximal random-effects often over-parameterize the model and a model simplification to a parsimonious structure is recommended \citep{bates_parsimonious_2015}.

\paragraph{Correlation analysis}
To better interpret the results, we considered whether the apparent positive effect of elaborated over verification feedback might be a confounding effect produced by a higher interest in the subject or by the better ability of the student. Examples of variables that could be associated with such confounding factors, apart from the use of feedback, are the average time spent on solving an exercise, the prior knowledge level, the ability of the student, the average response confidence, and the number of exercises solved in the platform. Thus, we examined whether the frequency of feedback positively correlated with some of the above mentioned variables computing the Pearson's correlation (\texttt{stats::cor.test(method = "pearson")} in R). Although we did not measure all possible confounding variables (e.g., quality of teaching, socio-economic status, or achievement in other subjects), the available data might offer insight into possible influences of student's motivation or ability in the effect of feedback.

\subsubsection{Unsupervised clustering of feedback use: RQ3}
To characterize students’ use of MER feedback, we analyzed behavior at two complementary temporal granularities: a time-aggregated view (overall per-student proportions of verbal/graphical/mathematical feedback) to capture what students tended to consult, and a time-resolved view (per-exercise sequences) to test how those choices evolved over time.

\paragraph{Time-aggregated MER-patterns}
To summarize preferred feedback strategies, we applied $k$-means clustering (in Python using \texttt{scikit-learn}'s \texttt{KMeans}) to per-student feature vector (\textit{verbal frequency}, \textit{graphical frequency}, \textit{ mathematical frequency}), the proportions of each feedback type among per student (variables defined in Section  \ref{subsect:analysis-plan}). The number of clusters $k$ was chosen by applying the elbow criteria to the silhouette metrics and the gap statistics \citep{rousseeuw_silhouettes_1987, thorndike_who_1953, tibshirani_estimating_2001}; details are in Appendix \ref{app:cumulative clusters}. We fitted the clusters separately for Cohorts~1 and~2 to check consistency (both Cohorts saw the same items and feedback), and we also refitted the model on the pooled sample to report a single, averaged solution for RQ3.

\paragraph{Time-resolved MER-patterns}
Because MER choices may change across exercises, we also clustered sequences of feedback selections (time-resolved MER patterns). For each student we formed a sequence over {\texttt{NF}, \texttt{V}, \texttt{G}, \texttt{M}, \texttt{VG}, \texttt{VM}, \texttt{GM}, \texttt{VGM}} as defined in Section~\ref{subsection: preprocessing}. To each feedback label we assigned a number \texttt{NF=0, V=1, G=2, M=3, VG=4, VM=5, GM=6, VGM=7}, in order to be able to calculate a degree of similarity between students. This encoding creates a scale in which smaller values indicate that fewer MERs were used. We computed pairwise distances between students’ sequences and applied hierarchical clustering (Ward linkage) to produce a dendrogram (in \textsf{R} with \texttt{stats::hclust(method = "ward.D2")} on Euclidean distances) (Appendix, Fig.~\ref{dendrogram}); from which the optimum number of clusters can be determined (Appendix, Fig.~\ref{elbow_long}). To keep sequence lengths comparable, we restricted this analysis to students who completed all 43 exercises (N=192 from both Cohorts combined, 29.5\% of eligible students); We combine both Cohorts for this part of the analysis, since we are not considering pre- and post-test results, but only feedback selection. 

\subsubsection{Group comparisons: RQ4-RQ6.}
\subsubsection{RQ4: association between MER-patterns and post-test performance}

To test whether post-test scores differed by feedback strategy, we analyzed the time-aggregated clusters from RQ3 but restricted the comparison to students who actually used elaborated feedback (clusters 1–3; cluster 0 was excluded). For each Cohort separately, we fitted an ANCOVA with post-test score (percent correct) as the outcome, feedback-strategy-cluster (three levels) as the factor, and pre-test score and feedback frequency ($p_\text{FB}$) as covariates; continuous predictors were mean-centered prior to estimation. We assessed the homogeneity-of-slopes assumption by adding the (feedback frequency)~$\!\times\!$~(feedback-strategy-cluster) interaction. When this interaction was not significant, we interpreted the cluster main effect and reported adjusted means with Bonferroni-adjusted pairwise contrasts; when the interaction was significant or borderline, we probed simple slopes of feedback frequency within each cluster and visualized adjusted means at the sample mean of feedback frequency. All tests were two-sided with $\alpha=.05$; we report $F$ statistics with degrees of freedom, $p$ values, adjusted mean differences (SE), and 95\% confidence intervals, calculated in R using \texttt{stats::aov}

\subsubsection{RQ5: Association between MER-patterns and initial representational competences}

To test whether students’ initial representational competences were related to their later MER-pattern choices, we compared competence levels across the time-aggregated feedback-strategy clusters from RQ3. Because this question concerns baseline abilities, the analysis was conducted only for Cohort~1 (the Cohort with the competence measures). We ran a one-way MANOVA with \emph{feedback-strategy cluster} (four levels, including the no-feedback group, Cluster~0) as the factor and the three competence scores (verbal, graphical, mathematical) as jointly analyzed dependent variables. We report Pillai’s trace as the omnibus statistic due to its robustness. If the multivariate test had been significant, we planned follow-up univariate ANOVAs with Bonferroni-adjusted pairwise contrasts; otherwise, no post-hoc testing was performed. For transparency, we also display cluster means with standard errors (Fig.~\ref{fig:init_comp}).

\subsubsection{RQ6: Do initial competence profiles moderate the effect of MER-feedback strategies?}

To examine whether associations between of MER-based feedback strategies and learning depends on students’ initial representational competences (Cohort~1), we first clustered students by their baseline competence profiles. We applied $k$-means to the three pre-test subscores (verbal, graphical, mathematical), selecting the number of clusters using the same criteria as in the feedback-use clustering (elbow, average silhouette, and gap statistic; Appendix~\ref{app:cumulative clusters}).

Next, we tested moderation with a two-way ANCOVA on post-test score. Fixed factors were (a) \emph{feedback-strategy cluster} from the time-aggregated analysis (three levels; students who never opened elaborated feedback, cluster~0, were excluded) and (b) \emph{competence cluster} (four levels). Pre-test score and frequency of elaborated-feedback use ($p_\text{FB}$) were entered as covariates; continuous predictors were mean-centered. We assessed the (feedback-strategy)~$\!\times\!$~(competence) interaction to test homogeneity of slopes across competence profiles. When the interaction was significant, we conducted Bonferroni-adjusted pairwise comparisons of feedback strategies \emph{within} each competence cluster and graphed adjusted means with 95\% confidence intervals. All tests were two-sided with $\alpha=.05$.

\section{Results and Discussion}
\label{sec: results}

\subsection{Linear mixed-effects regression: RQ1–RQ2}
\label{subsect:MLR}

We addressed RQ1–RQ2 with linear mixed-effects models estimated separately by Cohort, predicting post-test score from pretest score, platform performance, number of exercises solved, feedback frequency, and average response confidence, plus a centered interaction between platform performance and number of exercises solved. To account for classroom clustering we included a random intercept for lecture. Intraclass correlations were small (Cohort~1: ICC = .035; Cohort~2: ICC = .025), indicating that 3.5\% and 2.5\% of variance, respectively, was attributable to between-lecture differences.

Model summaries appear in Tables~\ref{tab:lmm_posttest_c1} and \ref{tab:lmm_posttest_c2}. Fixed effects explained a modest share of variance (marginal \(R^2 = .36\) for Cohort~1; \(R^2 = .49\) for Cohort~2), with total variance explained including random effects slightly higher (conditional \(R^2 = .38\) and \(R^2 = .50\), respectively).

Across Cohorts, pretest score and platform performance were the strongest predictors of post-test (Cohort~1: standardized \(\beta=0.30\) and \(0.36\); Cohort~2: \(\beta=0.41\) and \(0.39\)). The (platform)~$\!\times\!$~(exercises) interaction was positive and statistically significant in both Cohorts (Cohort~1: \(\beta=0.18\); Cohort~2: \(\beta=0.19\)), consistent with a larger association between platform score and post-test results for students who completed more exercises. This is expected given that platform scores become more reliable as the number of attempts increases.

The feedback frequency showed a small positive association with post-test performance in both Cohorts. This association was statistically significant in Cohort~1 (standardized \(\beta=0.13\), 95\%~CI [0.05, 0.22]) and positive but not conventionally significant in Cohort~2 (\(\beta=0.08\), 95\%~CI [0.00, 0.16], $p_{value} = .054$). Main effects of exercises solved and average confidence were small and not statistically significant after adjustment.

\begin{table}[!htbp]
  \centering
  \caption{Linear mixed-effects model predicting post-test score (Cohort~1, $n=351$, $J=34$ lectures)}
  \label{tab:lmm_posttest_c1}
  \begin{tabular*}{\textwidth}{@{\extracolsep{\fill}}lccc@{}}
    \toprule
    Predictor & Estimate $\pm$ SE & $t$ (df) & Std.\ $\beta$ [95\% CI] \\
    \midrule
    Intercept                                    & 23.85$^{***}$ $\pm$ 6.58 & 3.62 (329.0) & $-0.03$ [$-0.13,\,0.08$] \\
    Pre-test score                                & 0.335$^{***}$ $\pm$ 0.053 & 6.37 (346.4) & 0.30 [0.21, 0.39] \\
    Platform score (centered)                    & 0.411$^{***}$ $\pm$ 0.066 & 6.28 (339.3) & 0.36 [0.26, 0.47] \\
    Exercises solved (centered)                  & 0.077 $\pm$ 0.092         & 0.84 (147.5) & 0.06 [$-0.03, 0.15$] \\
    Feedback frequency             & 15.19$^{**}$ $\pm$ 5.03   & 3.02 (334.8) & 0.13 [0.05, 0.22] \\
    Average confidence                           & 2.60 $\pm$ 1.75           & 1.49 (341.1) & 0.07 [$-0.02, 0.17$] \\
    Platform $\times$ Exercises (centered)       & 0.017$^{***}$ $\pm$ 0.004 & 4.67 (343.2) & 0.18 [0.10, 0.26] \\
    \bottomrule
  \end{tabular*}
  \vspace{2pt}
  \begin{minipage}{\textwidth}
    \footnotesize
    \emph{Notes.} Linear mixed-effects model (REML) with random intercept for lecture ($J=34$).\\
    Random effects: SD$_{\text{lecture}}=4.03$ (Var$=16.27$), SD$_{\text{resid}}=21.03$ (Var$=442.34$); ICC $=0.035$.\\
    Model fit: Marginal $R^2=0.362$, Conditional $R^2=0.384$. Degrees of freedom via Satterthwaite.\\
    Continuous predictors were mean-centered; the interaction uses centered terms.\\
    Signif.\ codes: $^{***}p<.001$, $^{**}p<.01$, $^{*}p<.05$, $^{\cdot}p<.10$.
  \end{minipage}
\end{table}

\begin{table}[!htbp]
  \centering
  \caption{Linear mixed-effects model predicting post-test score (Cohort 2, $n=310$, $J=32$ lectures)}
  \label{tab:lmm_posttest_c2}
  \begin{tabular*}{\textwidth}{@{\extracolsep{\fill}}lccc@{}}
    \toprule
    Predictor & Estimate $\pm$ SE & $t$ (df) & Std.\ $\beta$ [95\% CI] \\
    \midrule
    Intercept                                    & 19.32$^{***}$ $\pm$ 5.13 & 3.77 (244.9) & $0.02$ [$-0.07,\,0.11$] \\
    Pre-test score                                & 0.48$^{***}$ $\pm$ 0.06  & 8.19 (292.5) & 0.41 [0.31, 0.51] \\
    Platform score (centered)                    & 0.51$^{***}$ $\pm$ 0.07  & 6.93 (291.9) & 0.39 [0.28, 0.51] \\
    Exercises solved (centered)                  & 0.14 $\pm$ 0.08          & 1.84 (42.7)  & 0.06 [$-0.03, 0.15$] \\
    Feedback frequency             & 9.21$^{\cdot}$ $\pm$ 4.78 & 1.93 (293.9) & 0.08 [0.00, 0.16] \\
    Average confidence                           & 2.61$^{\cdot}$ $\pm$ 1.40 & 1.86 (283.3) & 0.09 [$-0.01, 0.18$] \\
    Platform $\times$ Exercises (centered)       & 0.017$^{***}$ $\pm$ 0.004 & 4.66 (292.2) & 0.19 [0.11, 0.27] \\
    \bottomrule
  \end{tabular*}
  \vspace{2pt}
  \begin{minipage}{\textwidth}
    \footnotesize
    \emph{Notes.} Linear mixed-effects model (REML) with random intercept for lecture ($J=32$).\\
    Random effects: SD$_{\text{lecture}}=2.70$ (Var$=7.281$), SD$_{\text{resid}}=16.71$ (Var$=279.276$); ICC $=0.025$.\\
    Model fit: Marginal $R^2=0.491$, Conditional $R^2=0.504$. Degrees of freedom via Satterthwaite.\\
    Continuous predictors were mean-centered; the interaction uses centered terms.\\
    Signif.\ codes: $^{***}p<.001$, $^{**}p<.01$, $^{*}p<.05$, $^{\cdot}p<.10$.
  \end{minipage}
\end{table}

To investigate whether there were interaction effects between feedback frequency and prior knowledge, platform score or response certitude (RQ2), we conducted three additional multiple linear regression analysis, each including the respective interaction term: (1) feedback frequency and pre-test score, (2) feedback frequency and platform score, or (3) feedback frequency and average response confidence. The results are shown in Tables \ref{tab:lmm_interactions_c1} and \ref{tab:lmm_interactions_c2} for each Cohort. No significant interaction effects were found in either case for neither of the Cohorts.

\begin{table}[!htbp]
\centering
\caption{Interaction checks in linear mixed-effects models (Cohort~1; ML fits)}
\label{tab:lmm_interactions_c1}
\begin{tabular*}{\textwidth}{@{\extracolsep{\fill}}lcccc@{}}
\toprule
Interaction term & Std.\ $\beta$ [95\% CI] & $\Delta R^2_{\text{marg}}$ & LRT $\chi^2$(1) & $p$ \\
\midrule
pretest $\times$ feedback & -0.03 [-0.12, 0.07] & 0.00022 & 0.34 & .560 \\
platform $\times$ feedback & 0.05 [-0.02, 0.12] & 0.0029 & 1.71 & .191 \\
confidence $\times$ feedback & 0.03 [-0.05, 0.12] & 0.00088 & 0.58 & .446 \\
\bottomrule
\end{tabular*}
\vspace{2pt}
\footnotesize
\emph{Notes.} Reduced models included all main effects; full models added the listed interaction. Models were fit with ML (not REML) and compared via likelihood-ratio tests. $\Delta R^2_{\text{marg}}$ is the change in marginal $R^2$ (fixed effects only). Random effects were identical across models (random intercept for lecture). 
\end{table}

\begin{table}[!htbp]
\centering
\caption{Interaction checks in linear mixed-effects models (Cohort 2; ML fits)}
\label{tab:lmm_interactions_c2}
\begin{tabular*}{\textwidth}{@{\extracolsep{\fill}}lcccc@{}}
\toprule
Interaction term & Std.\ $\beta$ [95\% CI] & $\Delta R^2_{\text{marg}}$ & LRT $\chi^2$(1) & $p$ \\
\midrule
pretest $\times$ feedback & -0.03 [-0.12, 0.07] & 0.5$\cdot 10^{-5}$ & 0.8$\cdot 10^{-3}$ & .978 \\
platform $\times$ feedback & -0.06 [-0.13, 0.01] & 0.0041 & 2.74 & .098 \\
confidence $\times$ feedback & 0.02 | [-0.06, 0.09] & 0.00046 & 0.17 & .677 \\
\bottomrule
\end{tabular*}
\vspace{2pt}
\footnotesize
\emph{Notes.} Reduced models included all main effects; full models added the listed interaction. Models were fit with ML (not REML) and compared via likelihood-ratio tests. $\Delta R^2_{\text{marg}}$ is the change in marginal $R^2$ (fixed effects only). Random effects were identical across models (random intercept for lecture). 
\end{table}

Regarding possible confounding variables, in Table \ref{corr_mot_all}, we show the Pearson correlation coefficient between the feedback frequency and other motivation-related variables for both Cohorts. We found no correlation between the use of feedback and the pre-test score, the platform score, the number of exercises solved, or the average response confidence. This indicates that those students who spent more time reading feedback were not necessarily those who solved more exercises or had a higher prior knowledge or ability level or confidence in their response (potentially indicative of subject aptitude or interest). However, we found a moderate correlation between the use of feedback and the average time spent solving an exercise, indicating that students who spent more time reading feedback also spent more time previously thinking about the question asked.

\begin{table}[!htbp]
    \centering
    \caption{Pearson correlation of feedback frequency with other potentially motivation-related variables}
    \label{corr_mot_all}
\begin{tabular*}{\textwidth}{@{\extracolsep{\fill}}lll@{}}
    \toprule
        Variable & Cor. [95\% CI] (Cohort~1) &  Cor. [95\% CI] (Cohort 2) \\
        \midrule
        avg. response time & 0.43*** [0.34, 0.51] & 0.34*** [0.24, 0.44]\\
        pre-test score & 0.02 [-0.08, 0.12]& 0.05 [-0.06, 0.16]  \\
        platform score & -0.08 [-0.18, 0.02]& 0.08 [-0.03, 0.19] \\
        number of exercises solved & 0.03 [-0.13, 0.08]& -0.09 [-0.20, 0.02]  \\
        avg. confidence & 0.00 [-0.1, 0.1] & 0.10. [-0.01, 0.21] \\
    \bottomrule
    \end{tabular*}
    \vspace{2pt}
  \begin{minipage}{\textwidth}
    %\footnotesize
    \textit{Note.} $^{***} p<.001$, $^{**} p<.01$, $^{*} p<.05$, . p$< 0.1$
  \end{minipage}
\end{table}

In summary, the multiple linear regression analysis showed a small but significant positive association between the frequency with which students accessed elaborated feedback and their post-test scores in both cohorts (RQ1). In other words, students who opened elaborated feedback more frequently after answering tended to obtain higher post-test scores. This association remained independent of prior-knowledge, platform performance, and reported confidence for both Cohorts (RQ2). Taken together, these findings suggest that, in this context, access to elaborated feedback in addition to verification feedback is associated with higher post-test performance than access to verification feedback alone, across levels of prior mastery.

It is important to note that because our design was observational (no randomized groups), we cannot infer a causal relation between the frequency of use of elaborated
feedback and the learning gains. A confounding variable, like intrinsic or extrinsic motivation in the subject, could also cause both a higher frequency of use of elaborated feedback, as well as a higher post-test score, independently.

We argue that a motivated or interested student might also invest more time solving exercises, as well as solving more of them. The student might also show a higher prior-knowledge level, a higher performance during the training or a higher confidence. If motivation to perform well in the subject leads a student to read elaborated feedback more frequently and to achieve a higher score in the post-test, we may be able to infer this motivation by examining, for example, the amount of exercises that the student solves or the time taken to answer. However, we found only a moderate correlation between the frequency of feedback and the response time. If motivation is confounding the effect of feedback, that motivation does not cause students to solve more exercises on the platform, nor is it related to higher prior knowledge, platform performance, or confidence. An alternative explanation for the correlation between time spent on a question and frequency of feedback is that students who reflect longer on a question are also those who care more about the feedback. At this point it is important to note that some authors argue that, for feedback to affect learning, a certain amount of motivation must be present \citep{hattie_power_2007, depasque_effects_2015}. In order to learn from the feedback, students need to care about it. But caring about the feedback also means caring about the question.

\subsection{Unsupervised clustering to find MER-patterns: RQ3}
\label{subsec: clustering}
\subsubsection{Time-aggregated MER-patterns}
\label{subsubsetc:time-agg}

\begin{figure}
\centering
\includegraphics[width=13cm]{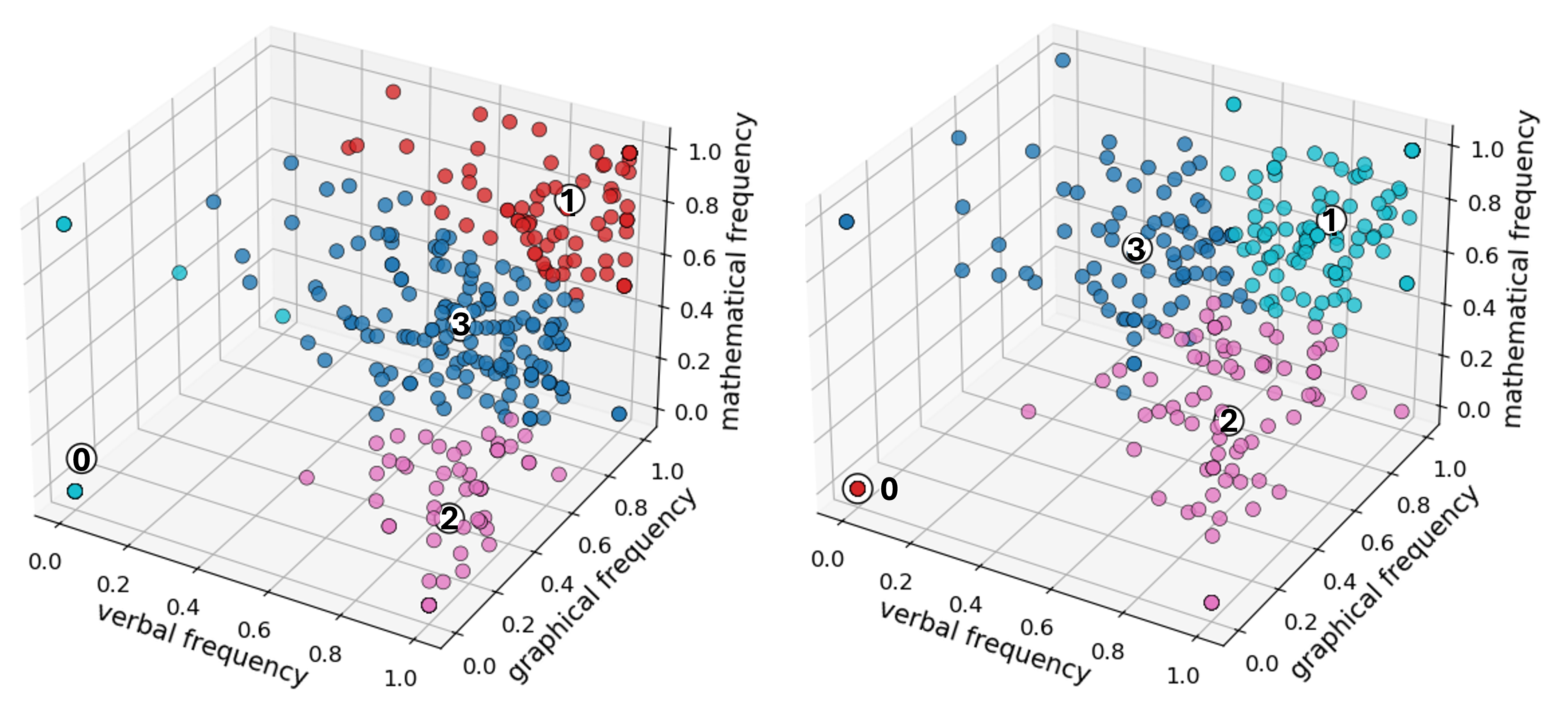}
\caption{\textbf{Student clusters generated using K-Means algorithm for Cohort~1 (left) and Cohort 2 (right).} Each dot represents a student according to the overall verbal, graphical, and mathematical frequency of feedback selected by the student. The clusters are differentiated by colors and their centroid is represented by the cluster labels $0, 1, 2, 3$.
}
\label{3d_clust_agg_2}
\end{figure}

In Fig.~\ref{3d_clust_agg_2}, we show the time-aggregated clusters found for Cohort~1 (left) and Cohort~2 (right). The left (right) graph of Fig.~\ref{3d_clust_agg_2} displays data points from Cohort~1~(2). The optimum number of clusters for both Cohorts was 4. The cluster centroids are represented in the graphs with cluster labels 0-3. In Table \ref{tab:clusters_agg}, we show the coordinates of the centroids of each cluster together with the number of students in that cluster for Cohort~1 (left graph) and Cohort~2 (right graph). Cluster~0 is centered, in both cases, at exactly 0\% feedback for the three categories, with 44 and 30 students for Cohort~1 and 2 respectively. Cluster~1 represents, both for Cohort~1 (101 students) and Cohort~2 (106) students, a high percentage of feedback (between 80\% and 91\%) in all categories. Cluster~2 is dominated by a higher percentage of verbal feedback in relation to graphical or mathematical feedback, especially for Cohort~1, with 67 and 79 students from each Cohort. Finally, Cluster~3 is centered at a moderate percentage of verbal, graphical, and mathematical feedback, although Cohort~1 (148 students) is again more biased towards verbal feedback than Cohort~2 (86 students).

\begin{table}[!htbp]
    \centering
        \caption{Centroids and sizes of time-aggregated clusters for MER-patterns for Cohort~1 (1) and 2 (2).}
        \label{tab:clusters_agg}
\begin{tabular*}{\textwidth}{@{\extracolsep{\fill}}lcccc@{}}
        \toprule
        Label &verbal freq. (1)/(2) &graphical freq. (1)/(2) &mathematical freq. (1)/(2) & Size (1)/(2) \\
        \midrule
        0 & 0.00/0.00 & 0.00/0.00 & 0.00/0.00 & 44/30 \\
        1 & 0.89/0.87 & 0.91/0.83 & 0.85/0.80 & 101/106 \\
        2 & 0.94/0.86 & 0.21/0.33 & 0.14/0.38 & 67/79  \\
        3 & 0.75/0.44 & 0.60/0.61 & 0.51/0.68 & 148/86  \\
        \bottomrule
    \end{tabular*}
    \vspace{2pt}
    \footnotesize\emph{Legend} Frequencies are proportions of representation use overal all feedback events.
\end{table}

\begin{figure}
\centering
\includegraphics[width=7.5cm]{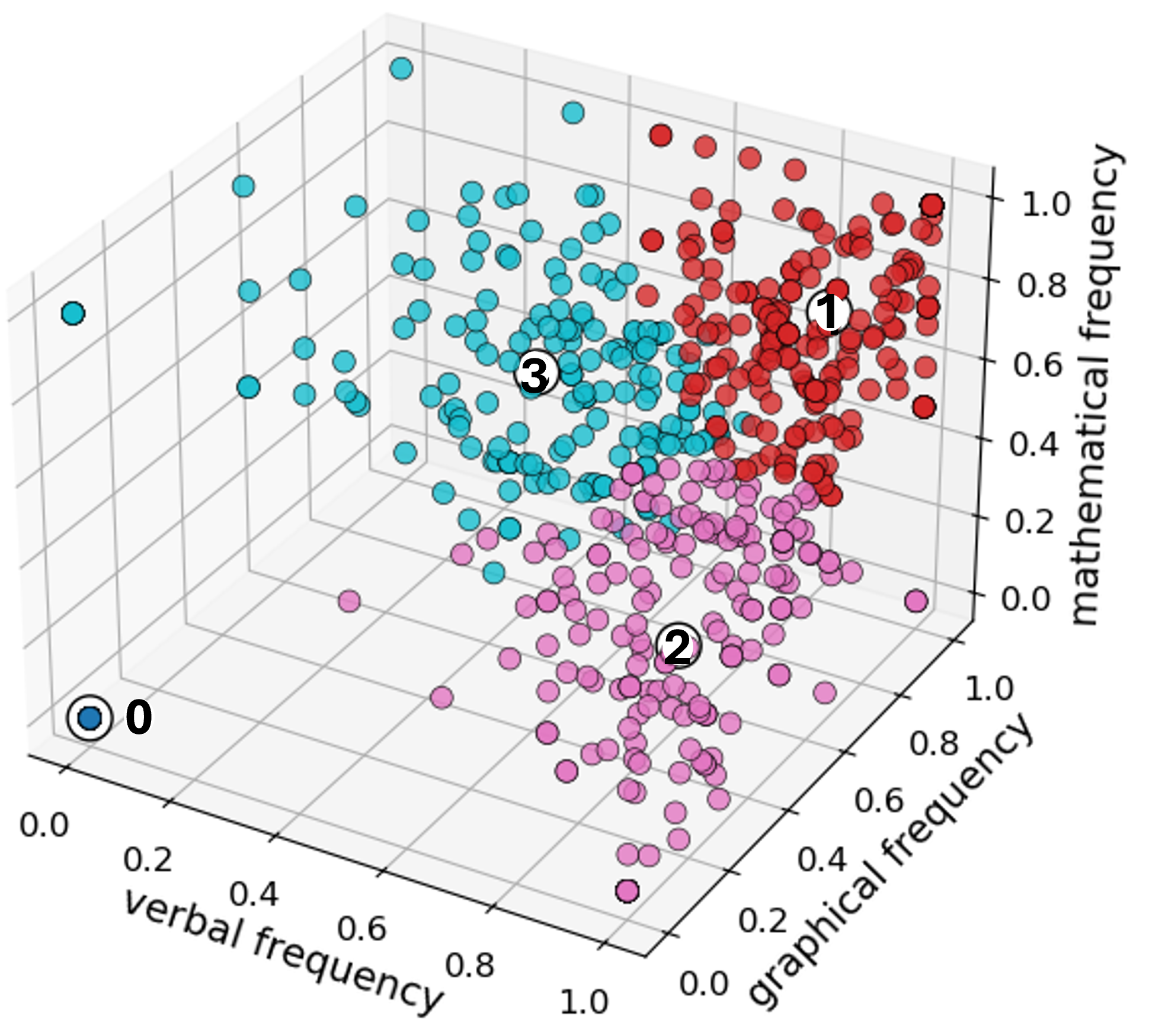}
\caption{\textbf{Student clusters generated using K-Means algorithm for Cohort~1 and Cohort 2 combined.} Each dot represents a student according to the overall verbal, graphical, and mathematical frequency of feedback selected by each student. The clusters are differentiated by colors and their centroid is tagged using labels $0, 1, 2, 3$.
}
\label{3d_clust_agg}
\end{figure}

In Fig.~\ref{3d_clust_agg}, we show all students from both Cohorts. Students belonging to different clusters are distinguished by color, with cluster centroids marked by labels ranging from 0 to 3. Table \ref{tab:clusters_agg_all} contains the coordinates of the different cluster-centroids and the size of each cluster, which are qualitatively similar to the centroids determined for each Cohort separately.

\begin{table}[!htbp]
    \centering
        \caption{Centroids and sizes of the time-aggregated clusters for MER-patterns for Cohorts 1 and 2 combined.}
        \label{tab:clusters_agg_all}
\begin{tabular*}{\textwidth}{@{\extracolsep{\fill}}lcccc@{}}
        \toprule
        Label & verbal frequency & graphical frequency & mathematical frequency & Size \\
        \midrule
        0 & 0.00 & 0.00 & 0.00 & 68 \\
        1 & 0.90 & 0.85 & 0.80 & 223 \\
        2 & 0.90 & 0.35 & 0.30 & 200  \\
        3 & 0.48 & 0.61 & 0.65 & 170  \\
        \bottomrule
    \end{tabular*}
    \vspace{2pt}
    \footnotesize\emph{Legend.} Frequencies are proportions of representation use overall all feedback events.
\end{table}

In summary, we observed three main MER-selection-strategies regarding elaborated feedback use (time-aggregated): consistently selecting all three MERs for every feedback consultation, evenly consulting the three MERs during the use of the platform but not simultaneously, or giving more weight to verbal feedback over the other types. A fourth group comprised students who never selected elaborated feedback and were exposed only to verification feedback. These tendencies were consistent across Cohorts.

\subsubsection{Time-resolved MER-patterns}

\begin{figure}[!htbp]
    \centering
    \includegraphics[width=\linewidth]{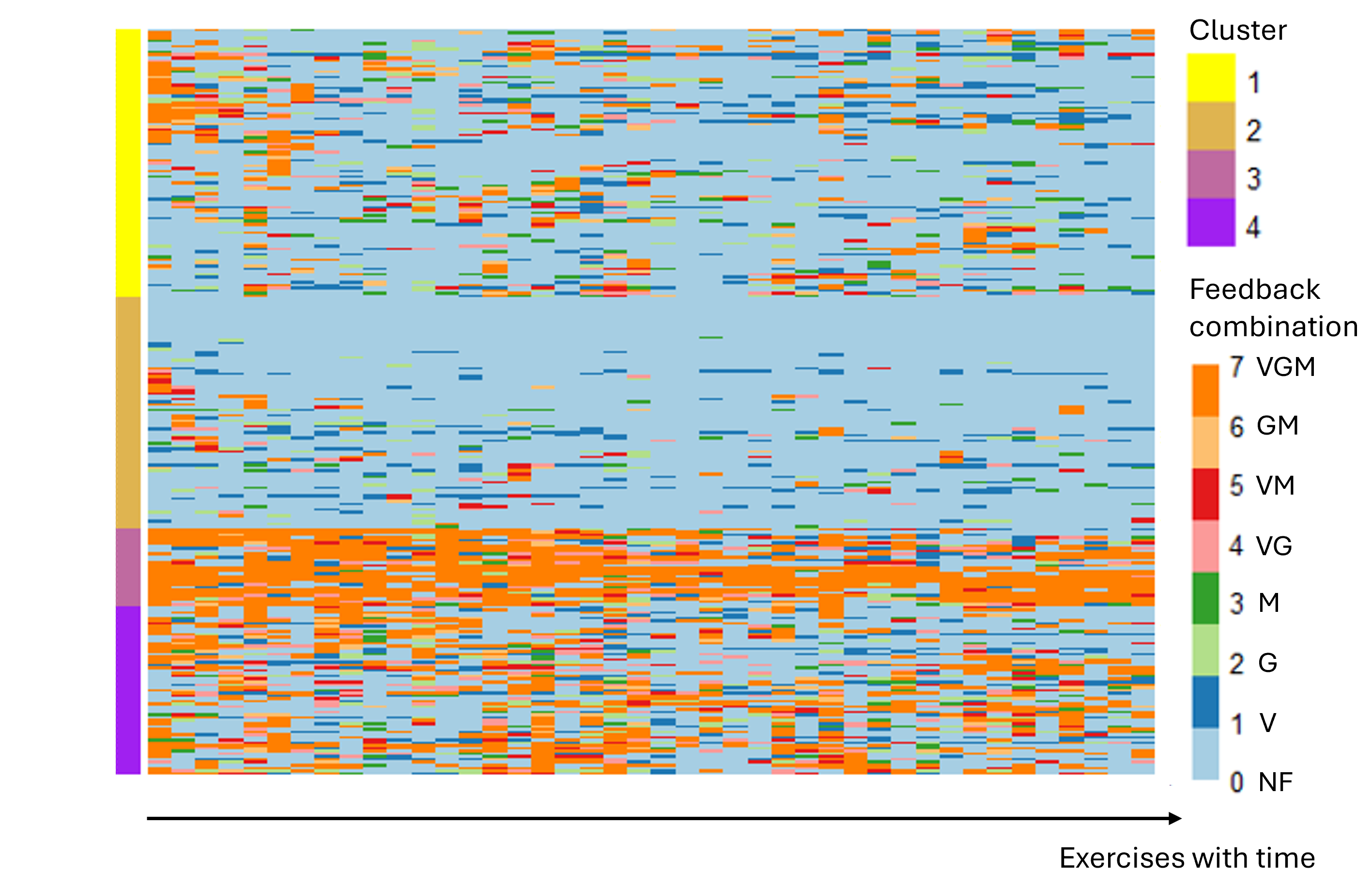}
    \caption{\textbf{A heatmap representation of the hierarchical clustering of students according to their time-resolved MER-patterns.} The color scale from 0 to 7 corresponds to the eight possible feedback combinations outlined in the main text (Section \ref{subsection: preprocessing}), and the color scale from 1 to 4 represents the four clusters found. In the horizontal dimension we represent the feedback combination per exercises over time, creating a feedback sequence for each student, vertically stacked.}
    \label{fig:heatmap}
\end{figure}

Fig.~\ref{fig:heatmap} shows the feedback sequences over time (horizontal axis and stacked vertically) of all students, and the respective calculated clusters, identified by a color bar on the left side of the figure. To represent the different feedback combinations, we use a heatmap. Each color of the heatmap scale represents a different combination of feedback. The cluster and feedback combination labels are specified to the right of the graph. The heatmap revealed that the most common feedback combination selected was the three representations simultaneously (14.7$\%$ of the heatmap). However, in the majority of attempted exercises, no feedback was selected (66.7$\%$ of the heatmap). Other feedback combinations occurrences ranged from 7.0$\%$ to 1.5\%. All occurrences are detailed in Table \ref{tab:occ_feedback}.

\begin{table}[!htbp]
  \centering
  \caption{Occurrence of MER-combinations in complete-exercise subset (n=192).}
  \label{tab:occ_feedback}
\begin{tabular*}{\textwidth}{@{\extracolsep{\fill}}lr@{}}
    \toprule
    MER combination & Occurrence (\%) \\
    \midrule
    No feedback                 & 66.74 \\
    Verbal                      & 7.04  \\
    Graphical                   & 2.78  \\
    Mathematical                & 2.19  \\
    Verbal and graphical        & 3.81  \\
    Verbal and mathematical     & 2.28  \\
    Graphical and mathematical  & 1.48  \\
    All representations         & 14.68 \\
    \bottomrule
  \end{tabular*}
  \vspace{2pt}
  \footnotesize\emph{Legend.} Percentages may not sum to 100 due to rounding.
\end{table}

By analyzing the clusters in the heatmap we observe that there is not enough variation in feedback selection over time to be able to identify groups with time differences in feedback combinations (80\% of all feedback selections is either no feedback or the three feedback types simultaneously). The characteristic that differentiates clusters seems to be the frequency with which feedback was selected, which is basically independent of the time dimension. This would mean that the time dimension does not add meaningful new information to our analysis.

Therefore, we conclude that our time-resolved MER-patterns analysis does not reveal distinct time-resolved MER-patterns. The sequences of feedback combinations over time are clustered based solely on feedback-selection frequency, rather than the types of feedback used. This happens because most of the students tended to either not select any feedback at all or to look at the three types of feedback simultaneously, and they do so consistently over time (although an overall decrease in the use of feedback towards the end of the experiment is observed for all clusters).

\begin{figure}[!htbp]
    \centering
    \includegraphics[width=8cm]{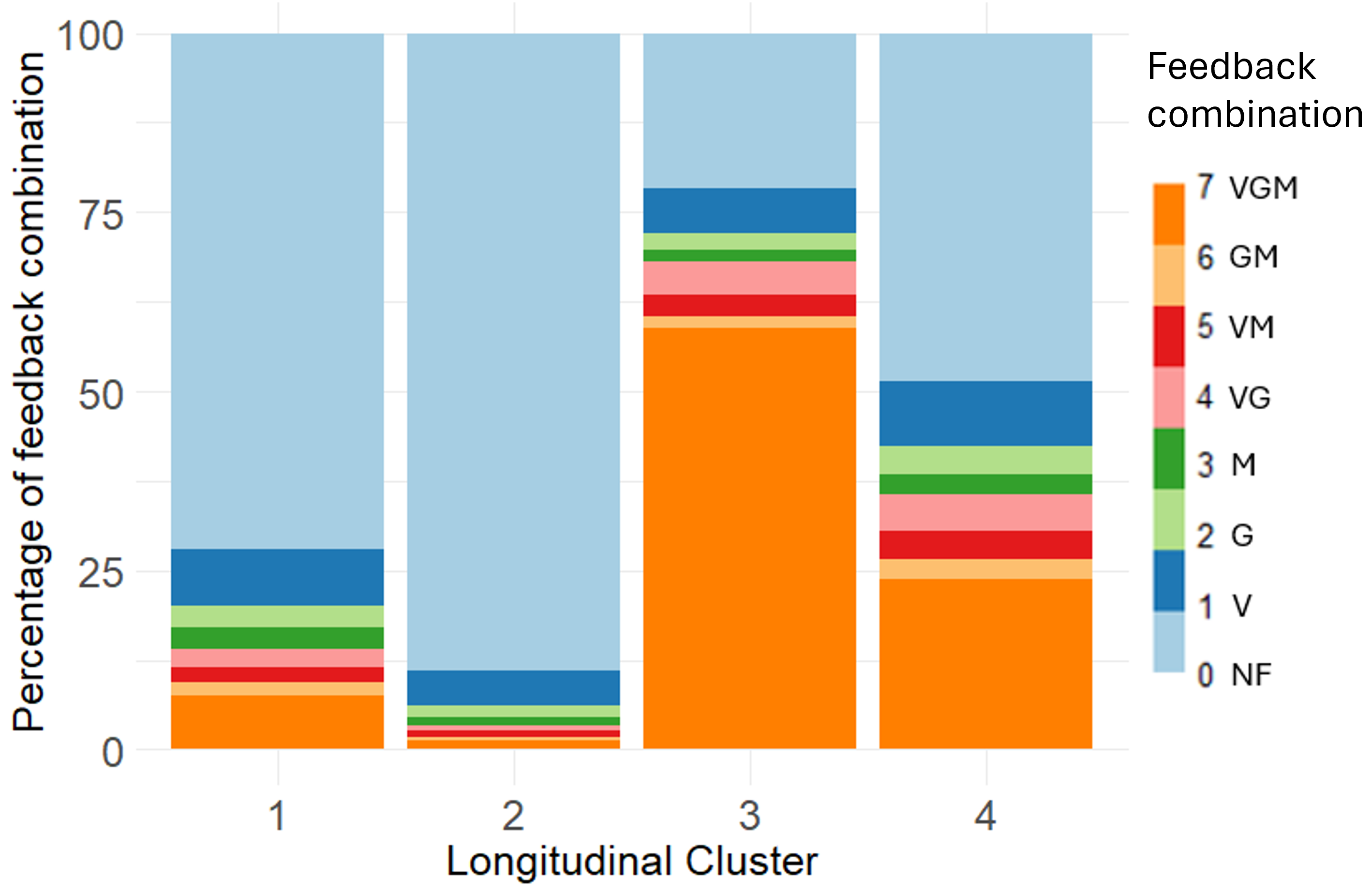}
    \caption{\textbf{Distribution of feedback combinations for each time-resolved cluster.} The categories "No feedback": 0, "verbal": 1, "graphical": 2, "mathematical": 3, "verbal and graphical": 4, "verbal and mathematical": 5, "graphical and mathematical": 6, "verbal, graphical and mathematical": 7, are represented in a color scale.}
    \label{fig:feed_dist}
\end{figure}

To facilitate interpretation of the clusters, and because the time dimension did not provide additional information, we removed this dimension and present the resulting distribution of feedback combinations in Fig.~\ref{fig:feed_dist}. We define these clusters as \textit{feedback-use} clusters. Cluster 1 has the highest proportion of no feedback, while Cluster 3 has the lowest. Cluster 3 is also characterized by a high proportion of the three MERs, followed by Clusters 4 and 2. The overall percentage of feedback selected is around 12\%, 29\%, 48\% and 77\% for clusters 1, 2, 3, and 4, respectively. 

In Fig.~\ref{fig:feed_dist}, it can be recognized that a higher use of feedback corresponds to a distinct MER-strategy, because the proportion of feedback combinations varies across feedback-use clusters. This means, in turn, that the distinct MER-strategy clusters found in section \ref{subsubsetc:time-agg} would present a significant variation on frequency of feedback use. In fact, an ANOVA test examining feedback frequency across feedback strategy clusters specifically reveals significant differences in the use of feedback between feedback-strategy groups in terms of frequency (Table \ref{tab:ancova_use}). A Tukey-HSD comparison shows that feedback-strategy-cluster 1 has significantly higher feedback use than feedback-strategy-cluster 2 and 3 (Table \ref{tab:tukey_pairs}). The mean feedback use per cluster is as follows: Cluster 1 $=0.374\pm0.0160$, Cluster 2 $=0.287\pm0.0134$, Cluster 3 = $0.258\pm0.0130$ clicks on elaborated feedback per total questions answered.

\begin{table}[!htbp]
  \centering
  \caption{ANCOVA of feedback frequency by strategy (combined Cohorts).}
  \label{tab:ancova_use}
  \begin{tabular*}{\textwidth}{@{\extracolsep{\fill}}lccccc@{}}
    \toprule
    ANCOVA term & Df & Sum Sq & Mean Sq & F & p \\
    \midrule
    Feedback Strategy Clusters & 2   & 1.469  & 0.7343 & 17.51 & $<$ .001*** \\
    Residuals                  & 590 & 24.743 & 0.0419 & —     & — \\
    \bottomrule
  \end{tabular*}
  \vspace{2pt}
  \footnotesize\textit{Note}: *** p$<$.001, ** p$<$.01, * p$<$.05.
\end{table}

\begin{table}[!htbp]
\centering
\caption{Tukey HSD pairwise comparisons for feedback frequency by feedback strategy cluster.}
\begin{tabular*}{\textwidth}{@{\extracolsep{\fill}}lcc@{}}
\toprule
Comparison & Mean diff [95\% CI] & {$p_{\text{adj}}$} \\
\midrule
Cluster 2 - Cluster 1 & -0.086184 [-0.133044, -0.039323] & $<$ .001*** \\
Cluster 3 - Cluster 1 & -0.115935 [-0.164927, -0.066944] & $<$ .001*** \\
Cluster 3 - Cluster 2 & -0.029752 [-0.079947,  0.020444] & .3454 \\
\bottomrule
\end{tabular*}
\vspace{2pt}
\footnotesize\textit{Note:} *** $p <$ 0.001, ** $p <$ 0.01, * $p <$ 0.05
\label{tab:tukey_pairs}
\end{table}

In conclusion, when exploring the time dimension, our clustering method could not distinguish between different strategies of MERs selection over time. It merely clustered students by how often they chose elaborated feedback (four groups in total, averaging around 12\%, 29\%, 48\% and 77\% of feedback frequency, respectively).

Interestingly, the results in Fig.~\ref{fig:feed_dist} show that students who selected elaborated feedback more often were also more likely to click all three types of feedback simultaneously. We can conclude that, when students were given the freedom to select elaborated feedback (in terms of frequency and format), those who requested feedback more frequently also tended to voluntarily check all the three MERs available. Indeed, Table \ref{tab:occ_feedback} shows that the most frequent combination of MERs was all three simultaneously (the second most frequent was verbal feedback, followed by verbal and graphical). In contrast, students who read elaborated feedback less often were more likely to select mainly verbal feedback, or to select the one or two MERs but not all of them simultaneously. Table \ref{tab:tukey_pairs} shows that ratio of feedback use by students in feedback-strategy-cluster 1 (mostly all three formats simultaneously) was significantly higher than that of students from feedback-strategy-cluster 2 (majority verbal) and 3 (all three formats but not simultaneously).

\subsection{Relation between MER-patterns, individual characteristics and performance: RQ4-RQ6}
\subsubsection{RQ4: how do the MER-patterns that student follow relate to their performance in the post-test?}

\begin{figure}[!htbp]
    \centering
    \includegraphics[width=11cm]{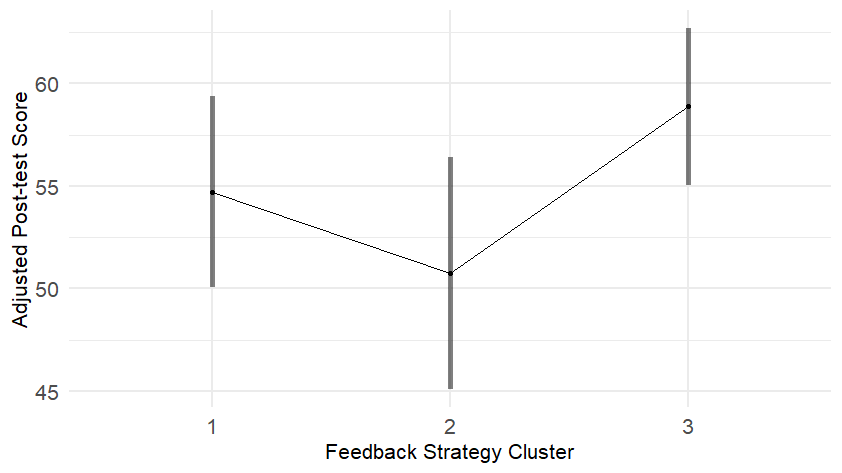}
    \caption{\textbf{Adjusted post-test score for each feedback-strategy-cluster from Cohort~1} (adjusted for pre-test and feedback frequency). Bars show 95\%confidence intervals.}
    \label{fig:pair-1}
\end{figure}

For Cohort~1, the (feedback frequency)~$\!\times\!$~(feedback-strategy-cluster) interaction was not significant, $F(2,309)=1.05$, $p=.350$. Like in Section \ref{subsect:MLR}, pre-test was a strong predictor of post-test performance, $F(1,309)=80.12$, $p<.001$, whereas the main effect of feedback frequency was not significant in this model, $F(1,309)=2.20$, $p=.139$. The feedback-strategy-cluster main effect was small and close to the conventional threshold, $F(2,309)=2.90$, $p=.057$. Bonferroni pairwise contrasts suggested only a trend for cluster~3 to outperform cluster~2 (diff$=-8.11$, SE$=3.47$, $t=-2.34$, $p=.060$). Adjusted means are shown in Fig.~\ref{fig:pair-1}; full ANCOVA and contrasts appear in Tables~\ref{tab:ancova_cumu_1} and~\ref{tab:posthoc-feedback-bayern}.

\begin{table}[!htbp]
    \centering
    \caption{ANCOVA for post-test score of Cohort~1 across time-aggregated clusters excluding cluster 0.}
    \label{tab:ancova_cumu_1}
\begin{tabular*}{\textwidth}{@{\extracolsep{\fill}}lccccc@{}}
        \toprule
        ANCOVA Results & DF & Sum Sq. & Mean Sq. & F-Value & p-value \\
\midrule
    Pre-test Score & 1 & 44231 & 44231 & 80.123 & $<$ 2e-16 *** \\
    Feedback Frequency & 1 & 1215 & 1215 & 2.201 & .1390 \\
    Feedback Strategy Clusters & 2 & 3201 & 1600 & 2.899 & .0566 . \\
    Frequency $\times$ Strategy & 2 & 1162 & 581 & 1.053 & .3503 \\
    Residuals & 309 & 170583 & 552 & -- & -- \\
\bottomrule
    \end{tabular*}
    \vspace{2pt}
    \footnotesize\textit{Note:} *** $p <$ 0.001, ** $p <$ 0.01, * $p <$ 0.05, . p$< 0.1$
\end{table}

\begin{table}[!htbp]
    \centering
    \caption{Pairwise contrasts of adjusted post-test means by feedback strategy clusters (Cohort~1)}
    \label{tab:posthoc-feedback-bayern}
    \begin{tabular*}{\textwidth}{@{\extracolsep{\fill}}lcccc@{}}
        \toprule
        Contrast & Estimate & SE & $t$-ratio & $p$-value \\
        \midrule
        Feedback Cluster 1 -- 2 & 3.95 & 3.74 & 1.055 & .8771 \\
        Feedback Cluster 1 -- 3 & -4.16 & 3.08 & -1.350 & .5338 \\
        Feedback Cluster 2 -- 3 & -8.11 & 3.47 & -2.340 & .0597 . \\
        \bottomrule
    \end{tabular*}
    \vspace{2pt}
    \footnotesize\emph{Legend:} differences are adjusted after applying the linear regression model summarized in Table \ref{tab:ancova_cumu_1}. \textit{Note:} *** $p <$ 0.001, ** $p <$ 0.01, * $p <$ 0.05, . p$< 0.1$
\end{table}

\begin{figure}[!htbp]
    \centering
    \includegraphics[width=11cm]{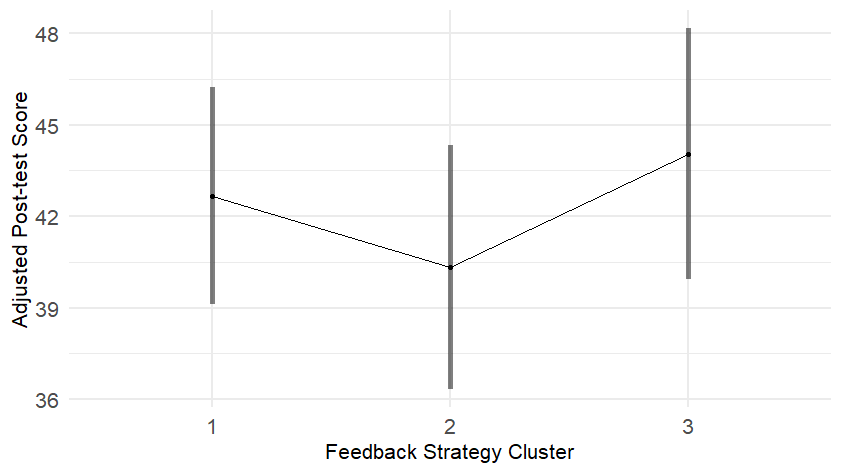}
    \caption{\textbf{Adjusted post-test score for each feedback-strategy-cluster from Cohort 2 for a fixed feedback frequency of 0.30.} The dots represent the mean post-test adjusted score for the cluster, whereas the bars represent 95\% confidence intervals.}
    \label{fig:pair-2}
\end{figure}

For Cohort~1, the (feedback frequency)~$\!\times\!$~(feedback-strategy-cluster) term bordered significance, $F(2,264)=2.97$, $p=.053$, so we examined simple slopes. Only cluster~3 showed a significant positive association between feedback frequency and post-test score (estimate$=33.83$, SE$=13.02$, $t=2.60$, $p=.010$); clusters~1 and~2 did not (Table~\ref{tab:simple-slopes-hessen}). At the sample mean (feedback frequency) = 0.30, adjusted means did not differ, $F(2,264)=0.41$, $p=.667$ (Fig.~\ref{fig:pair-2}).

\begin{table}[!htbp]
    \centering
    \caption{ANCOVA for post-test score of Cohort 2 across time-aggregated clusters excluding cluster 0.}
    \label{tab:anova_cumu_2}
     \begin{tabular*}{\textwidth}{@{\extracolsep{\fill}}lccccc@{}}
        \toprule
        ANCOVA Results & DF & Sum Sq. & Mean Sq. & F-Value & p-value \\
        \midrule
    Pre-test Score & 1 & 59006 & 59006 & 181.332 & $<$ 2e-16 *** \\
    Feedback Frequency & 1 & 1288 & 1288 & 3.957 & .0477 * \\
    Feedback Strategy Clusters & 2 & 264 & 132 & 0.406 & .6665 \\
    Frequency $\times$ Strategy & 2 & 1935 & 967 & 2.973 & .0529 . \\
    Residuals & 264 & 85906 & 325 & -- & -- \\
    \bottomrule
    \end{tabular*}
    \vspace{2pt}
    \footnotesize \textit{Note:} *** $p <$ 0.001, ** $p <$ 0.01, * $p <$ 0.05, . p$< 0.1$
\end{table}

\begin{table}[!htbp]
    \centering
    \caption{Conditional effects of feedback frequency on post-test performance for each feedback strategy cluster.}
    \label{tab:simple-slopes-hessen}
    \begin{tabular*}{\textwidth}{@{\extracolsep{\fill}}lcccc@{}}
        \toprule
        Feedback Strategy Cluster & Estimate & SE & $t$-value & $p$-value \\
        \midrule
        Cluster 1 & -0.76 & 7.86 & -0.10 & .92 \\
        Cluster 2 & 19.17 & 11.31 & 1.70 & .09 . \\
        Cluster 3 & 33.83 & 13.02 & 2.60 & .01 * \\
        \bottomrule
    \end{tabular*}
    \vspace{2pt}
    \footnotesize\emph{Legend:} The linear slopes indicate the extent to which feedback frequency predicts post-test scores within clusters, after accounting for pre-test performance and feedback frequency.\textit{Note:} *** $p <$ 0.001, ** $p <$ 0.01, * $p <$ 0.05, . p$< 0.1$
\end{table}

Taken together, we did not observe robust mean differences between feedback strategies after covariate adjustment in either Cohort. However, the pattern is consistent across Cohorts with a possible advantage for the strategy that alternates across the three representations (cluster~3): in Cohort~1 this appears as a near-threshold contrast with the verbal-dominant strategy (cluster~2), and in Cohort~2 as a clear positive slope linking greater use of elaborated feedback to higher posttest scores only within cluster~3.

\subsubsection{RQ5: how do the MER-patterns that student follow relate to their initial representational competences?}

\begin{figure}[!htbp]
    \centering
    \includegraphics[width=11cm]{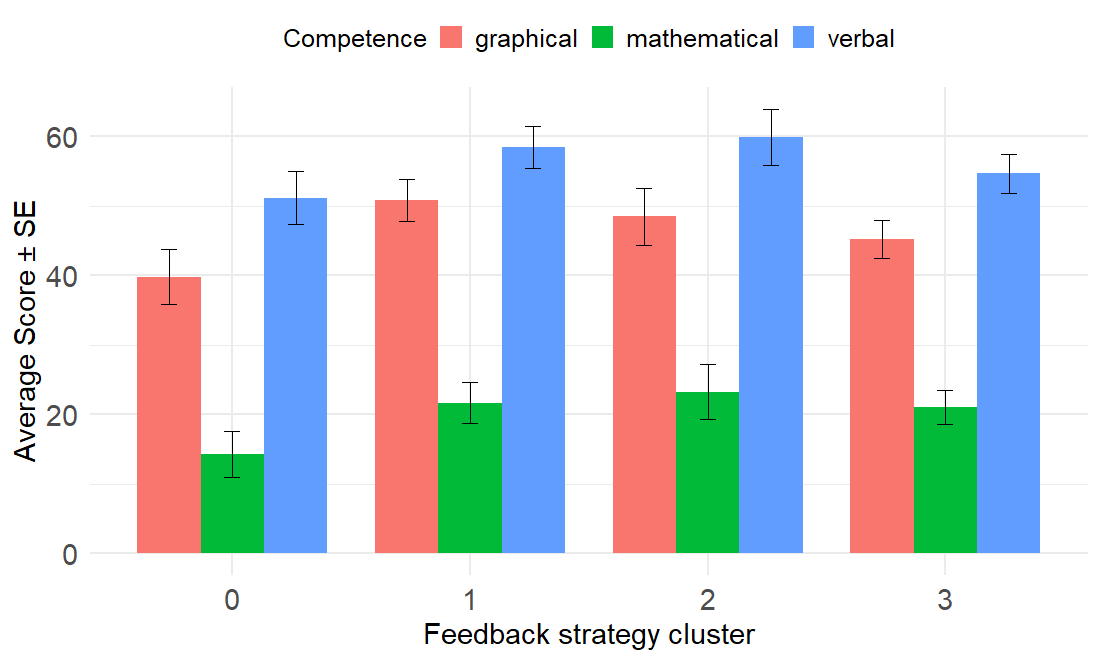}
    \caption{\textbf{Average initial representational competence level for each competence and each feedback strategy cluster.} The error bars represent standard errors.}
    \label{fig:init_comp}
\end{figure}

For Cohort~1, the multivariate test showed no evidence of differences in initial competences across feedback-strategy-clusters (Pillai’s trace $=0.018$, $F=0.74$, $p=.67$; Table~\ref{tab:manova}). The plot of cluster means (Fig.~\ref{fig:init_comp}) illustrates this: average verbal, graphical, and mathematical competences are broadly similar across clusters. Descriptively, students in Cluster~0 (no elaborated-feedback use) tended to score lowest on all three scales, but these differences were not statistically significant. In other words, within this cohort, we cannot attribute learners’ preference for one combination of MERs over another to their initial representational competences.

\begin{table}[!htbp]
    \centering
    \caption{MANOVA for initial representational competences between different feedback strategy clusters for Cohort~1. }
    \label{tab:manova}
    \begin{tabular*}{\textwidth}{@{\extracolsep{\fill}}lcccc@{}}
        \toprule
        MANOVA Results & DF & Pillai & F-Value & p-value \\
        \midrule
        Feedback Strategy Cluster & 3 & 0.018 & 0.74 & .67 \\
        Residuals & 356 &  &  &  \\
        \midrule
    \end{tabular*}
\end{table}

\subsubsection{RQ6: Do initial competence profiles moderate the effect of feedback strategies?}

Clustering the three pretest competence scores yielded four distinct baseline profiles (Table~\ref{tab:clusters_comp}): a low-all cluster (0), a high-all cluster (1), a profile with relatively strong verbal and graphical but weak mathematical competence (2), and a profile with strong verbal but weak graphical and mathematical competence (3).

\begin{table}[!htbp]
    \centering
        \caption{Centroids and sizes of the initial representational
competence clusters.}
\label{tab:clusters_comp}
\begin{tabular*}{\textwidth}{@{\extracolsep{\fill}}lcccc@{}}
        \toprule
        Label & verb. comp. score &  graph. comp. score &  math. comp. score & Size \\
        \midrule
        0 & 20.65 & 18.63 & 12.12 & 118 \\
        1 & 80.35 & 68.94 & 76.97 & 52 \\
        2 & 69.32 & 74.73 & 12.15 & 124  \\
        3 & 71.10 & 21.45 & 7.84 & 66  \\
        \bottomrule
    \end{tabular*}
     \vspace{2pt}
    \footnotesize\emph{Abbreviations:} comp. = competences; verb. = verbal; graph. = graphical; math. = mathematical.
\end{table}

\begin{figure}[!htbp]
    \centering
    \includegraphics[width=12cm]{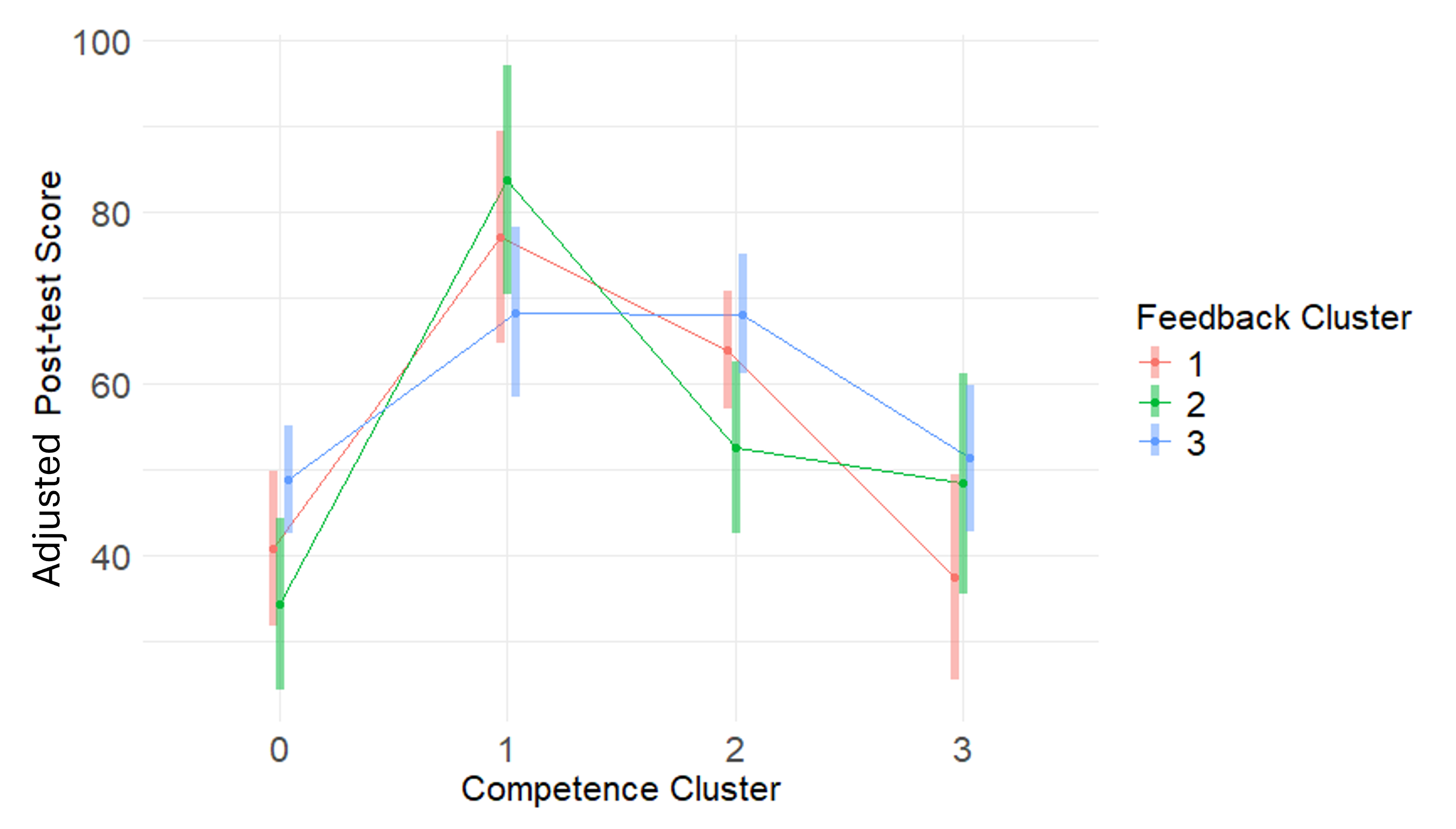}
    \caption{\textbf{Interaction graph between feedback-strategy and competence clusters for Cohort~1.} The x-axis represents  competence clusters according to the labels described in Table \ref{tab:clusters_comp}. Additionally, within each competence cluster, we distinguish by color the adjusted post-test score for each feedback strategy cluster. The legend numbers refer to the labels described in Table \ref{tab:ancova_cumu_1}. The bars represent 95\% confidence intervals.}
    \label{fig:int-two-clust}
\end{figure}

The two-way ANCOVA indicated a significant (feedback-strategy)~$\!\times\!$~(competence) interaction, $F=2.33$, $p=.033$, alongside a main effect of competence cluster (Table~\ref{tab:anova-two-way}). We therefore probed feedback-strategy differences within each competence profile (Table~\ref{tab:posthoc-feedback-within-comp}; Fig.~\ref{fig:int-two-clust}). In the clusters characterized by lower mathematical skill or uniformly lower competences (clusters~2 and~0), students who alternated across all three representations (feedback-strategy~3) achieved higher adjusted posttest scores than those who relied primarily on verbal feedback (strategy~2), with Bonferroni-adjusted contrasts reaching or approaching significance. In contrast, these strategy differences were not detectable within the mixed profile with stronger competencies (cluster~3), and the trend reversed in the highest-competence group (cluster~1), where the verbal-dominant strategy (2) tended to perform as well as or better than the alternating strategy (3), although differences were not statistically significant.

\begin{table}[!htbp]
    \centering
    \caption{ANCOVA for post-test score of Cohort~1 across time-aggregated clusters excluding cluster 0.}
    \label{tab:anova-two-way}
\begin{tabular*}{\textwidth}{@{\extracolsep{\fill}}lccccc@{}}
        \toprule
        Two-way ANCOVA Results & DF & Sum Sq. & Mean Sq. & F-Value & p-value \\
        \midrule
        Feedback Strategy Cl. (Factor A) & 2 & 1899 & 949 & 1.733 & .1785 \\
        Competence cluster (Factor B) & 3 & 43759 & 14586 & 26.621 & $10^{-15}$ *** \\
        Feedback frequency (covariate) & 1 & 1055 & 1055 & 1.925 & .1663 \\
        Factor A $\times$ Factor B (int.) & 6 & 7654 & 1276 & 2.328 & .0326 * \\
        Residuals & 303 & 166025 & 548 & -- & -- \\
        \bottomrule
    \end{tabular*}
     \vspace{2pt}
    \footnotesize \textit{Note:} *** $p <$ 0.001, ** $p <$ 0.01, * $p <$ 0.05, . p$< 0.1$
\end{table}

\begin{table}[!htbp]
    \centering
    \caption{Pairwise comparisons of feedback-strategy-clusters within each competence cluster (Bonferroni-adjusted).}
    \label{tab:posthoc-feedback-within-comp}
\begin{tabular*}{\textwidth}{@{\extracolsep{\fill}}lccccc@{}}
        \toprule
        Competence Cluster & Estimate & SE & $t$-ratio & $p$-value \\
        \midrule
        \multicolumn{5}{l}{\textbf{Competence cluster = 0}} \\
        Feedback Cluster 1 -- 2 & 6.52 & 6.87 & 0.949 & 1.0000 \\
        Feedback Cluster 1 -- 3 & -8.01 & 5.59 & -1.431 & .4602 \\
        Feedback Cluster 2 -- 3 & -14.53 & 6.04 & -2.407 & .0501 . \\
        \midrule
        \multicolumn{5}{l}{\textbf{Competence cluster = 1}} \\
        Feedback Cluster 1 -- 2 & -6.64 & 9.21 & -0.721 & 1.0000 \\
        Feedback Cluster 1 -- 3 & 8.67 & 8.04 & 1.079 & .8440 \\
        Feedback Cluster 2 -- 3 & 15.31 & 8.44 & 1.815 & .2116 \\
        \midrule
        \multicolumn{5}{l}{\textbf{Competence cluster = 2}} \\
        Feedback Cluster 1 -- 2 & 11.41 & 6.21 & 1.835 & .2023 \\
        Feedback Cluster 1 -- 3 & -4.15 & 4.98 & -0.834 & 1.0000 \\
        Feedback Cluster 2 -- 3 & -15.56 & 6.21 & -2.504 & .0385 * \\
        \midrule
        \multicolumn{5}{l}{\textbf{Competence cluster = 3}} \\
        Feedback Cluster 1 -- 2 & -10.90 & 8.90 & -1.225 & .6648 \\
        Feedback Cluster 1 -- 3 & -13.80 & 7.45 & -1.852 & .1948 \\
        Feedback Cluster 2 -- 3 & -2.90 & 7.82 & -0.371 & 1.0000 \\
        \bottomrule
    \end{tabular*}
    \vspace{2pt}
    \footnotesize \textit{Note:} *** $p <$ 0.001, ** $p <$ 0.01, * $p <$ 0.05, . p$< 0.1$
\end{table}

Taken together, our results support a competence-contingent effect of feedback strategy. For students with weaker representational foundations, particularly those with lower mathematical competence, alternating across verbal, graphical, and mathematical feedback was associated with higher adjusted post-test performance than relying primarily on a single (verbal) format. This pattern is consistent with prior work on the representational dilemma, which recommends exposing learners to not-yet-mastered representations in a paced way to avoid overload \citep{rau_supporting_2017,bollen_student_2017}. By contrast, for students with stronger initial representational competence, additional representational formats did not yield detectable gains, and the trend in the highest-competence group favored a more minimal, verbal-dominant approach. Although those differences were not statistically significant within that subgroup, the direction aligns with expertise-reversal accounts \citep{tetzlaff_cornerstone_2025}: as competences consolidate, extra representational elaboration can add redundancy rather than information, reducing efficiency.

Overall, these findings suggest a pragmatic guideline: when competence is low, encourage alternating use of all three representations (rather than only verbal or all three at once), whereas when competence is high, concise verbal feedback may suffice. We emphasize that these are observational associations; the competence-by-strategy interaction warrants confirmation in preregistered experiments that manipulate both representation format and dosage.

\subsection{Limitations}
Our study was an observational cohort conducted in natural classroom conditions, so the associations we report should not be interpreted causally; unmeasured confounding (e.g. motivation, teacher practices) may remain. Selection processes at the school/teacher level and attrition within classes may also bias estimates: not all eligible students used the platform or completed both tests. Information bias is possible because exposures were derived from log events that record opening/selection of feedback rather than depth of processing; logs can also contain technical noise (e.g., brief or accidental clicks), which may misclassify feedback strategy and frequency.

\section{Conclusions}

Across the two upper-secondary cohorts, more frequent consultation of elaborated feedback was associated with higher post-test performance, regardless of students’ prior knowledge, ability, or confidence. Students’ feedback-seeking clustered into distinct MER-strategies, and the benefit of a given strategy depended on initial representational competence. Learners with weaker representational foundations tended to benefit more from alternating across verbal, graphical, and mathematical feedback, whereas for learners with stronger representational competence additional formats offered no detectable advantage and may be even counterproductive.

Our results contribute to the understanding of adaptive feedback and the use of MERs in physics learning, and underscore the importance of providing flexible feedback options that can be tailored to each student's cognitive and metacognitive profiles. As digital platforms continue to evolve, integrating adaptive feedback mechanisms that account for both learner characteristics and representational dynamics promises more personalized and effective physics instruction. With the emergence of LLMs and their text-heavy instruction, it becomes particularly important to remain aware of the benefits that visual or mathematical instruction can bring, especially for those students who still lack a strong representational basis. MLLMs could potentially mitigate this limitation if trained and prompted appropriately \citep{bewersdorff_taking_2025}. 

Future research should further validate these findings through experimental designs that can disentangle causal effects of feedback elaboration from motivational or engagement-related factors. Investigating the temporal dynamics of representational use could clarify how students develop and refine their strategies across time. Moreover, integrating MLLMs into adaptive tutoring systems is a promising path to personalize both the content and format of feedback according to learners’ evolving competences, enabling real-time adaptation of feedback complexity and representational load, thereby optimizing learning.

\section*{Appendix}
\appendix
\renewcommand{\thefigure}{A.\arabic{figure}}
\renewcommand{\thetable}{A.\arabic{table}}
\setcounter{figure}{0}
\setcounter{table}{0}
\section{Attrition analysis across cohorts}\label{app:attrition}
To assess whether attrition was systematic within each cohort, we compared the pre-test scores of students who completed both the pre- and post-tests (“finishers”) with those who completed only the pre-test (“non-finishers”). The sample sizes for attrition analysis are shown in Table \ref{tab:attrition_sizes}.

Cohort 1 consisted of 423 students, of whom 351 (83\%) completed both tests and 72 did not. Pre-test score distributions were mildly right-skewed in both groups (skewness: finishers = 0.37; non-finishers = 0.40). Welch’s t-test showed that finishers scored significantly higher than non-finishers (Table \ref{tab:attrition_stats}). A Mann–Whitney U test confirmed this difference, U = 14 319.5, p = .024. The effect size was small-to-moderate (Cohen’s d = 0.31), indicating a modest but systematic tendency for higher-performing students to remain the study.

Cohort 2 included 925 students, of whom 310 (34\%) completed both tests and 615 did not. Pre-test score distributions were more strongly skewed (skewness: finishers = 1.20; non-finishers = 1.33). In contrast to Cohort 1, finishers and non-finishers did not differ significantly in pre-test performance. Welch’s t-test indicated no difference (Table \ref{tab:attrition_stats}), and the Mann–Whitney U test yielded a similar result, U = 100 622.5, p = .31. The effect size was negligible (Cohen’s d = 0.03), indicating no detectable attrition bias within this cohort (Table).

Taken together, these analyses indicate that attrition was systematic in Cohort 1 cohort but not in Cohort 2. While this does not invalidate the findings based on finishers, it call for extra caution when generalizing the results.

\begin{table}[!htbp]
  \centering
  \caption{Sample sizes for attrition analysis by cohort}
  \label{tab:attrition_sizes}
  \begin{tabular*}{\textwidth}{@{\extracolsep{\fill}}lccc@{}}
    \toprule
    Cohort & Finished ($n$) & Not finished ($n$) & Attrition rate \\
    \midrule
    1 & 351 & 72  & 17.0\% \\
    2 & 310 & 615 & 66.5\% \\
    \bottomrule
  \end{tabular*}
  \vspace{2pt}
  \begin{minipage}{\textwidth}
    \footnotesize
    \emph{Notes.}  
    Attrition rate is computed as the proportion of students who completed the pre-test but did not complete the post-test.
  \end{minipage}
\end{table}

\begin{table}[!htbp]
  \centering
  \caption{Comparison of pre-test scores between finishers and non-finishers by cohort}
  \label{tab:attrition_stats}
  \begin{tabular*}{\textwidth}{@{\extracolsep{\fill}}lcccccc@{}}
    \toprule
    Cohort & Skew$_\text{fin}$ & Skew$_\text{non}$ & $t$ (Welch) & $p$-value & U-test $p$ & Cohen's $d$ \\
    \midrule
    1 & 0.37 & 0.40 & 2.40 & .018 & .024 & 0.31 \\
    2 & 1.20 & 1.33 & 0.51 & .609 & .306 & 0.03 \\
    \bottomrule
  \end{tabular*}
  \vspace{2pt}
  \begin{minipage}{\textwidth}
    \footnotesize
    \emph{Notes.}  
    Skewness values describe the distribution of pre-test scores in each group.  
    $t$-values correspond to Welch's unequal-variance $t$-test.  
    U-test values refer to the Mann--Whitney U test (two-sided).  
    Cohen's $d$ is computed using the pooled standard deviation.  
    All analyses use pre-test scores only; post-test scores are unavailable for non-finishers.
  \end{minipage}
\end{table}

\renewcommand{\thefigure}{B.\arabic{figure}}
\renewcommand{\thetable}{B.\arabic{table}}
\setcounter{figure}{0}
\setcounter{table}{0}
\section{Psychometrics of pre-test, post-test and Intervention Items}\label{app:psychometrics}

In this section, we show psychometric analyses for the pre-, post-test and platform exercises for Cohorts 1 and 2.

In Table \ref{tab:test_statistics}, psychometric data for pre-test 1 (used with Cohort~1) and pre-test 2 (used with Cohort 2) is presented. The test statistics for both tests show reasonable values of discrimination index, point biserial coefficient, reliability index and Ferguson´s delta. All these metrics scored above the minimum threshold found in the literature for both tests. Only the difficulty index is slightly below the recommended values for both tests. This was expected, as students were not yet expected to possess enough knowledge to answer the pre-test correctly at the beginning of the study. The values were calculated using the formulas from \citep{ding_approaches_2009}. The psychometric data scores higher in reliability and discrimination in the case of our self-designed test (Test 1), while the test made out of selected parts from validated knowledge concept test seems to be less robust in this aspect, although still above the recommended threshold. Therefore, we conclude that both our self-designed test and the test built using validated concept tests are adequate for the study.

\begin{table}[!htbp]
\centering
\caption{Test statistics for the pre-test compared with desired values according to \citep{ding_approaches_2009}.}
\label{tab:test_statistics}
\begin{tabular*}{\textwidth}{@{\extracolsep{\fill}}lccccc@{}}
        \toprule
Pre-test Statistics & Pre-test 1 & Pre-test 2 & Desired Values \\
\midrule
Avg. Difficulty Index              & 0.27 & 0.26 & $\left[0.30,\, 0.90\right]$ \\
Avg. Discrimination Index          & 0.63 & 0.53 &  $\geq$ 0.30 \\
Avg. Point Biserial Coefficient    & 0.53 & 0.49 &  $\geq$ 0.20 \\
Reliability Index             & 1.16    & 0.78        & $\geq$ 0.70 \\
Ferguson’s Delta              & 0.95    & 0.94        & $\geq$ 0.90 \\
\bottomrule
\end{tabular*}
\end{table}

In Table \ref{tab:test_statistics_post}, psychometric data for post-test 1 (used with Cohort~1) and post-test 2 (used with Cohort 2) is presented. The test statistics for both tests show reasonable values of discrimination index, point biserial coefficient, reliability index and Ferguson´s delta. All of these metrics score above the minimum threshold found in the literature for both tests. We conclude that both our self-designed test (Test 1) and the test built using validated concept tests (Test 2) are sufficiently robust for the study.

\begin{table}[!htbp]
\centering
\caption{Test statistics for the post-test compared with desired values according to \protect\citep{ding_approaches_2009}.}
\label{tab:test_statistics_post}
\begin{tabular*}{\textwidth}{@{\extracolsep{\fill}}lccccc@{}}
        \toprule
Post-test Statistics & Post-test 1 & Post-test 2 & Desired Values \\
\midrule
Avg. Difficulty Index              & 0.46 & 0.42 & $\left[0.30,\, 0.90\right]$ \\
Avg. Discrimination Index          & 0.66 & 0.54 &  $\geq$ 0.30 \\
Avg. Point Biserial Coefficient    & 0.49 & 0.49 &  $\geq$ 0.20 \\
Reliability Index             & 0.77    & 0.79        & $\geq$ 0.70 \\
Ferguson’s Delta              & 0.97    & 0.98        & $\geq$ 0.90 \\
\bottomrule
\end{tabular*}
\end{table}

\begin{figure}[!htbp]
    \centering
    \includegraphics[width=10cm]{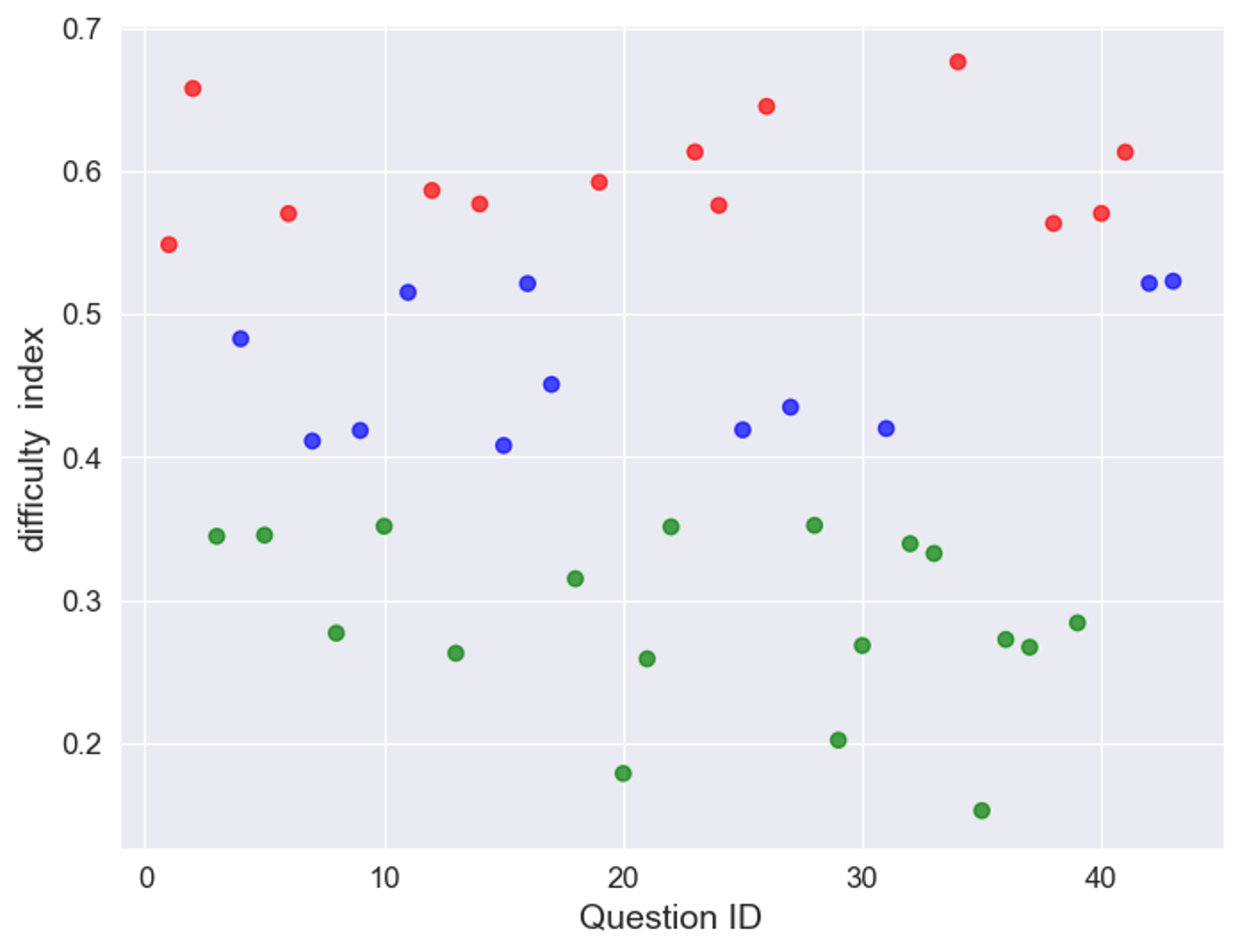}
    \caption{\textbf{Difficulty index for the 43 exercises used in the platform KI4SCool.} Color-coded are three clusters identified using k-means clustering (green: difficult, blue: medium, red: easy).}
    \label{fig:ratis_dist}
\end{figure}

Lastly, we conducted item-level psychometric analysis of the platform exercises. We are especially interested in determining the difficulty level of each exercise, as this could affect the effectiveness of elaborated feedback \citep{hattie_power_2007, shute_focus_2008, chaiklin_zone_2003}. On the one hand, elaborated feedback tends to be less useful if the exercises are too easy, as the questions being covered are not suitable to close possible knowledge gaps in students \citep{hattie_power_2007}. On the other hand, too difficult items would situate students out of their zone of proximal development \citep{chaiklin_zone_2003}. In Table \ref{tab:item_statistics} the results of the item analysis are shown. Additionally, we present in Fig.~\ref{fig:ratis_dist} a cluster analysis for the difficulty index of each exercise. There are 18 exercises labeled as \textit{difficult} (green dots), 12 as \textit{medium} (blue dots) and 13 as \textit{easy} (red dots). The average difficulty is 0.47. Since the distribution was centered around 0.47, and there were few extreme values (only 10 exercises might be too difficult, difficulty\_index$<$0.3), and no exercises that were too easy (difficulty\_index$>$0.7), we conclude that the exercises have a suitable difficulty level to facilitate learning from elaborated feedback.

\begin{table}[!htbp]
\centering
\caption{Item statistics for the intervention exercises compared with desired values according to \protect\citep{ding_approaches_2009}.}
\label{tab:item_statistics}
\begin{tabular*}{\textwidth}{@{\extracolsep{\fill}}lccccc@{}}
\toprule
Item Statistics & Values & Desired Values \\
\midrule
Difficulty Index              & Average of 0.47 & $\left[0.30,\, 0.90\right]$ \\
Discrimination Index          & Average of 0.98 & $\geq$ 0.30 \\
Point Biserial Coefficient    & Average of 0.55 & $\geq$ 0.20 \\

\bottomrule
\end{tabular*}
\end{table}

\renewcommand{\thefigure}{C.\arabic{figure}}
\renewcommand{\thetable}{C.\arabic{table}}
\setcounter{figure}{0}
\setcounter{table}{0}
\section{Finding the optimum number of clusters}
In this section we show the visual methods used to determine the optimum number of clusters for both the K-Mean algorithm and the hierarchical clustering.
 
\subsection{Time-aggregated Feedback Clusters and Competence Clusters}\label{app:cumulative clusters}

\begin{figure}[!htbp]
    \centering
    \includegraphics[width=12cm,height=14.5cm]{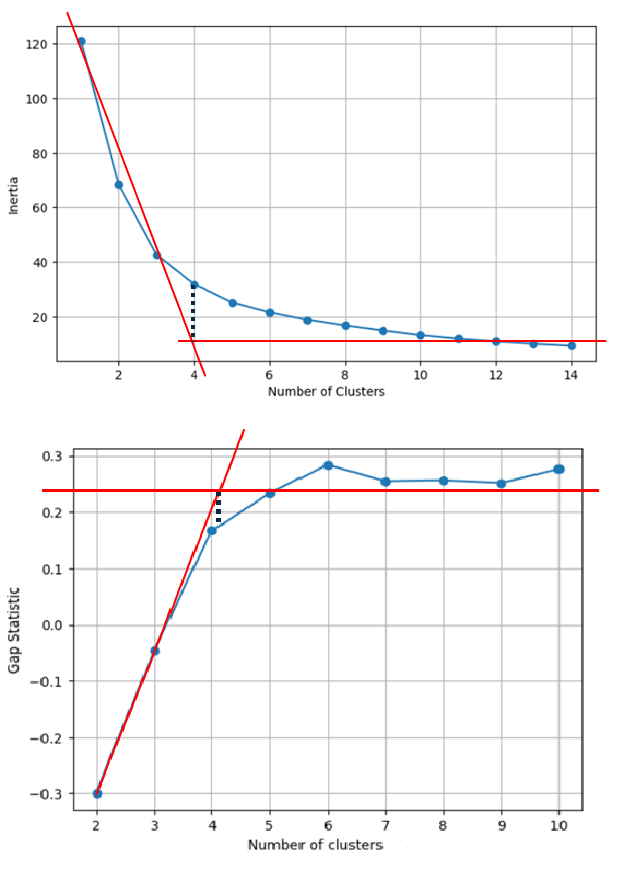}
    \caption{\textbf{Inertia (upper graph) and gap statistics (lower graph) for the number of time-aggregated feedback-strategy-clusters.} The red lines represent the elbow method that is applied to determine the optimum number of clusters.}
    \label{fig:num_cl_mer}
\end{figure}

\begin{figure}[!htbp]
    \centering
    \includegraphics[width=12cm,height=14.5cm]{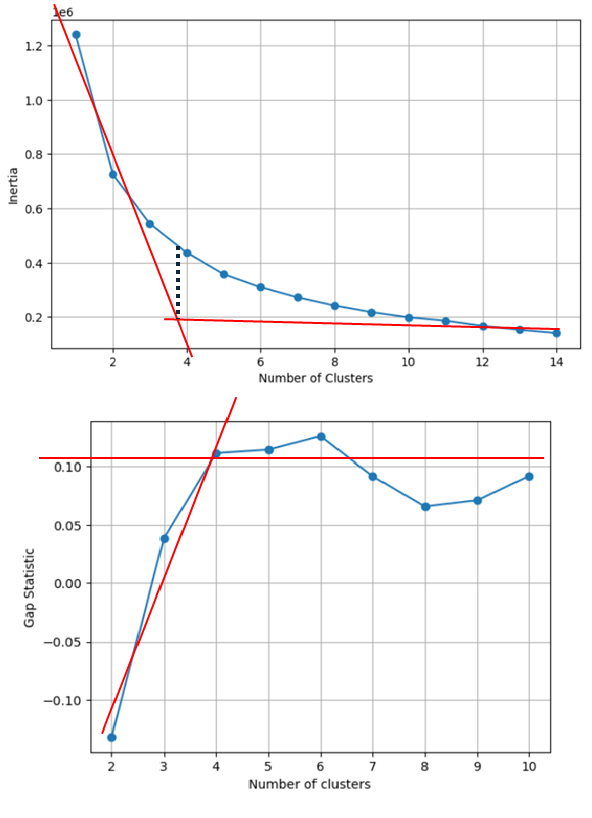}
    \caption{\textbf{Inertia (upper graph) and gap statistics (lower graph) for the number of clusters of initial representational competence.} The red lines represent the elbow method that is applied to determine the optimum number of clusters.}
    \label{fig:num_cl_comp}
\end{figure}

To identify the optimal number of clusters when using the K-Means algorithm, we applied the elbow method to two different metrics: the inertia and the gap statistics \citep{thorndike_who_1953, rousseeuw_silhouettes_1987, tibshirani_estimating_2001}. In Fig.~\ref{fig:num_cl_mer} both methods are illustrated for clusters of time-aggregated feedback strategy, whereas in Fig.~\ref{fig:num_cl_comp} the methods are applied to the competence clusters. In both cases and for both methods we concluded that the optimum number of clusters is four.

\subsection{Time-resolved Feedback Clusters}
\label{app: longitudinal clusters}
\begin{figure}[!htbp]
\centering
\includegraphics[width=12cm]{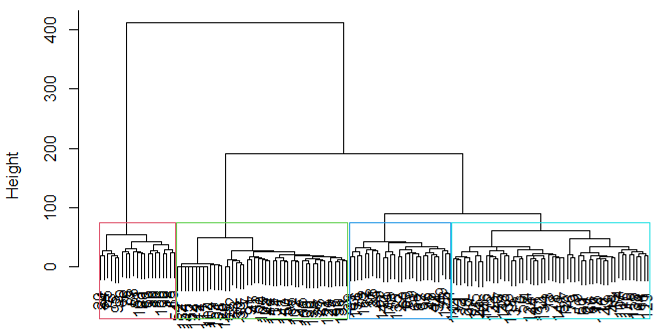}
\caption{\textbf{Student clusters generated using hierarchical clustering for time-resolved MER-patterns.} The colored lines enclose the four optimum clusters.
}
\label{dendrogram}
\end{figure}
\begin{figure}[!htbp]
\centering
\includegraphics[width=12cm,height=7cm]{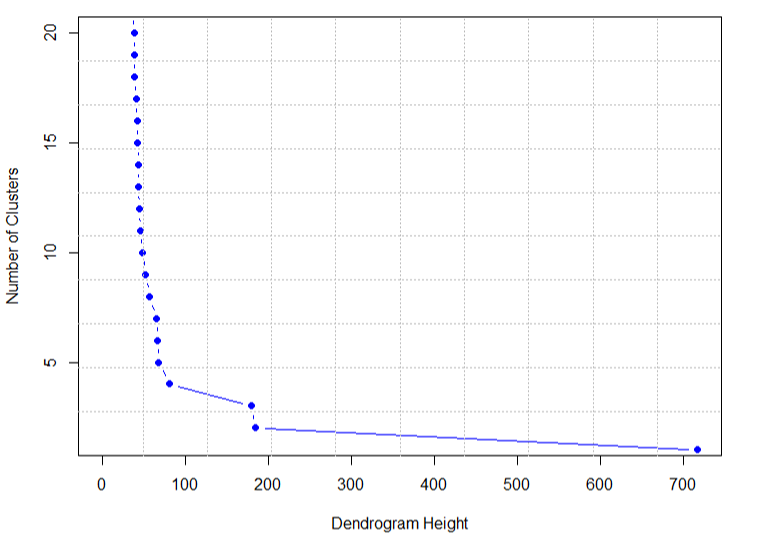}
\caption{\textbf{Number of clusters as a function of the dendrogram height.} An inflection point in the graph happens at number of clusters = 4.
}
\label{elbow_long}
\end{figure}

In Fig.~\ref{dendrogram} we show the dendrogram generated by the computed hierarchical clustering. At the bottom of the dendrogram, each student forms an individual cluster. The first two more similar students in terms of time-resolved MER-patterns are joined to form the first cluster. The magnitude \textit{Height} represents the degree of dissimilarity between clusters. Examining at the variation of height with each additional cluster (Fig.~\ref{elbow_long}, we observed a notable decrease after four clusters. Therefore, we conclude that the optimum number of clusters is four.

\bibliography{references}
\end{document}